\documentclass[apj,twocolumn]{aastex62}
\usepackage{lineno}
\usepackage{gensymb}
\usepackage{lineno}
\usepackage{amsmath}
\usepackage{wasysym}
\usepackage{float}
\usepackage{multirow}
 \usepackage{graphicx}
\usepackage{natbib}

\graphicspath{{./}{figures/}}
\NewPageAfterKeywords
\begin{document}
\title{\large Thermodynamic and Energetic Limits on Continental Silicate Weathering Strongly Impact the Climate and Habitability of Wet, Rocky Worlds}

\author{R.J. Graham}
\affil{University of Oxford}
\author{Ray Pierrehumbert}
\affil{University of Oxford}
\correspondingauthor{R.J. Graham}
\email{robert.graham@physics.ox.ac.uk}


\begin{abstract}
The ``liquid water habitable zone'' (HZ) concept is predicated on the ability of the silicate weathering feedback to stabilize climate across a wide range of instellations. However, representations of silicate weathering used in current estimates of the effective outer edge of the HZ do not account for the thermodynamic limit on concentration of weathering products in runoff set by clay precipitation, nor for the energetic limit on precipitation set by planetary instellation. We find that when the thermodynamic limit is included in an idealized coupled climate/weathering model, steady-state planetary climate loses sensitivity to silicate dissolution kinetics, becoming sensitive to temperature primarily through the effect of temperature on runoff and to $\mathrm{pCO_2}$ through an effect on solute concentration mediated by pH.  This increases sensitivity to land fraction, $\mathrm{CO_2}$ outgassing, and geological factors such as soil age and lithology, all of which are found to have a profound effect on the position of the effective outer edge of the HZ.  The interplay between runoff sensitivity and the energetic limit on precipitation leads to novel warm states in the outer reaches of the HZ, owing to the decoupling of temperature and precipitation. We discuss strategies for detecting the signature of silicate weathering feedback through exoplanet observations in light of insights derived from the revised picture of weathering.

\end{abstract}

\section{Introduction} \label{sec:intro}
The classical ``liquid water habitable zone'' (HZ) is the orbital region around stars where rocky planets with N$_2$-H$_2$O-CO$_2$ atmospheres can potentially support stable liquid water on their surfaces \citep{huang1959occurrence,hart1979habitable,Kasting93,Kopparapu:2013,ramirez2018more}. The inner edge of the HZ is the instellation at which a planet's oceans are likely to undergo rapid escape to space, either due to enhanced loss of water from a warm, wet stratosphere (the ``moist greenhouse state'') or due to full ocean evaporation and subsequent escape (the ``runaway greenhouse state'') \citep[e.g.][]{Kasting88, Kasting93, Kopparapu:2013}. The outer edge of the HZ is the instellation at which a planet cannot maintain a temperature above the freezing point of water anywhere on its surface with an atmosphere of N$_2$, H$_2$O, and CO$_2$ alone \citep[e.g.][]{kadoya2019outer}. Depending on conditions like the spectrum of the host star, the outer edge may be defined by the ``maximum greenhouse limit'', where an increase to the CO$_2$ partial pressure on a hypothetical planet would increase the Rayleigh scattering albedo enough to outweigh the greenhouse effect and cause the planet to freeze, or it may be defined by the ``CO$_2$ condensation limit'', where CO$_2$ can no longer accumulate to higher levels in the atmosphere because it is forced to condense or deposit at the planet's surface \citep[e.g.][]{Kasting93, Kopparapu:2013fp}.

The HZ concept as currently understood relies on the existence of a negative feedback mechanism on ocean-bearing rocky planets to control the concentration of atmospheric CO$_2$ and maintain surface liquid water at a variety of instellations. Without some kind of long-term negative feedback on climate, the question of whether a given planet's atmospheric CO$_2$ level is consistent with the presence of surface liquid H$_2$O at that planet's particular instellation is largely a matter of luck, and the lifetime of habitability for those planets lucky enough to start off temperate will be relatively short, due to the gradual brightening of stars. \citet{Walker-Hays-Kasting-1981:negative} (WHAK) suggested that temperature-, runoff-, and pCO$_2$-dependent weathering of silicate minerals (a process which produces cations that react with oceanic carbonate ions to form carbonate minerals that are sequestered in Earth's crust and subsequently recycled through the mantle) has served the role of stabilizing feedback on Earth, drawing down CO$_2$ as the sun has brightened from $\sim$70$\%$ of its current luminosity over the past 4 billion years (Ga) \citep{SAGAN:1972p1233,bahcall2001solar}. \citet{kasting1988climate} hypothesized that this silicate weathering feedback also operates on rocky ocean- and land-bearing exoplanets, and this idea underpins much subsequent work regarding the HZ (though see e.g. \citet{Pierrehumbert:2011p3366,kite2018habitability,ramirez2018ice} for non-standard habitability scenarios that do not involve the silicate weathering feedback). Recently, studies modeling planetary weathering behavior have suggested that enhanced weathering due to high pCO$_2$ in the outer reaches of the habitable zone may act to draw CO$_2$ down to levels that force a planet into the snowball state at instellations greater than the ``maximum greenhouse limit'' or ``CO$_2$ condensation limit'', effectively decreasing the width of the HZ, but the magnitude (and even the existence) of this effect is strongly dependent on the details of the silicate weathering feedback \citep{Menou2015,haqq2016limit,abbot-2016,kadoya2019outer}. Accurately representing silicate weathering in models is therefore necessary for predicting the extent of the HZ and the behavior of planets within it. These issues will become increasingly crucial as observation and analysis of temperate rocky planets and their atmospheres by future telescopes becomes commonplace. Here, we focus on continental weathering, though a full evaluation of habitability would, among other things, require a treatment of seafloor weathering as well (see Section \ref{subsubsec:limitations} for discussion of this issue).

In this study, we use a simple 0-dimensional climate model and the weathering framework developed in \citet{maher2014hydrologic} (MAC) to re-examine planetary weathering behavior and the outer edge of the habitable zone. The MAC model of weathering is the first to include a thermodynamic limit on cation concentration in runoff due to equilibration between dissolving silicates and precipitating clays. We also include an energetic limit on precipitation and runoff that accounts for the fact that evaporation (and therefore precipitation) is ultimately driven by instellation, which implies the existence of a maximum precipitation rate for a planet, defined as the rate at which all incoming stellar radiation must be used for latent heat of evaporation \citep{Pierrehumbert-2002:hydrologic, o2008hydrological}. We find that including these limits in global weathering models strongly impacts the climate and weathering behavior of wet, rocky exoplanets in the HZ. In particular, including the thermodynamic limit on solute concentration increases the importance of hydrology and surface properties like land fraction and soil age for global weathering fluxes, alternately expanding or contracting the effective width of the habitable zone depending on choice of parameters. With the energetic limit on precipitation, global precipitation ceases to be a function of temperature under some circumstances, leading to fundamental changes in the functioning of the silicate weathering feedback.  

In the remainder of this paper, we will more thoroughly introduce the silicate weathering feedback, along with its thermodynamic and energetic limits, which have not been included in previous studies of exoplanet habitability (Section \ref{sec:weathering}); describe the simple models we used to examine the impact of these limits on the behavior of planets within the HZ (Section \ref{sec:methods}); present the results of our calculations (Section \ref{sec:results}); and discuss implications of these results (Section \ref{sec:discussion}).

\section{The Continental Silicate Weathering Feedback} \label{sec:weathering}
The continental silicate weathering feedback is a mechanism proposed by WHAK to explain how Earth has maintained a relatively stable, temperate climate over geologic time despite volcanic outgassing of CO$_2$ and the long-term brightening of the Sun. In Earth's carbonate-silicate cycle \citep[e.g.][]{siever1968sedimentological, Walker-Hays-Kasting-1981:negative}, volcanic CO$_2$ is released into the atmosphere/ocean system by the metamorphism of carbonate minerals in the planet's interior. The CO$_2$ acts as an acid in aqueous silicate weathering reactions that speed up the release of Ca$^{2+}$ and Mg$^{2+}$ cations which are then carried by rivers to the oceans, where they react with carbonate ions and precipitate as carbonate minerals. These minerals sink to the ocean floor, where they are subducted into the Earth, completing the cycle. This complex process can be represented schematically as:
\begin{linenomath}
\begin{align}
    \text{CaSiO}_3 + \text{CO}_2 \leftrightarrow \text{CaCO}_3+\text{SiO}_2
\end{align}
\end{linenomath}
where the rightward direction represents a silicate mineral reacting with (and consuming) a CO$_2$ molecule to form a carbonate mineral and silica (SiO$_2$), and the leftward direction represents an overall reaction where a carbonate mineral is metamorphosed with silica in Earth's mantle to form a silicate and CO$_2$. In actuality, a variety of silicate minerals can take the place of CaSiO$_3$ in the above equation, and secondary minerals, e.g. clays, precipitate in regions being weathered and play vital roles in controlling the rates of the weathering reactions \citep{alekseyev1997change,maher2009role, maher2014hydrologic}, with important implications to be discussed in Section \ref{subsec:mac}.

\subsection{The WHAK Model}\label{subsec:whak}
To understand where a negative feedback enters into this process in the WHAK model, we can isolate the step where the silicate mineral is dissolved:
\begin{linenomath}
\begin{align}
\begin{split}
    \label{eqn:wollastonite}\text{CaSiO}_3 (s) + 2\text{CO}_2 (g) + \text{H}_2\text{O} (l) \xrightarrow{} \\\text{Ca}^{2+} (aq) + 2\text{HCO}_3^- (aq) + \text{SiO}_2 (aq)
\end{split}
\end{align}
\end{linenomath}
Laboratory studies demonstrate that the kinetics of silicate dissolution depend directly on temperature and pH for a variety of silicate minerals \citep[e.g.][]{schott1985dissolution, brady1991effect,knauss1993diopside,oxburgh1994mechanism,welch1996feldspar,chen1998diopside,weissbart2000wollastonite,oelkers2001experimental,palandri2004compilation,carroll2005dependence,golubev2005experimental,bandstra2008data,brantley2008kinetics}. Atmospheric pCO$_2$ influences temperature through its greenhouse effect and pH through its action as a weak acid in aqueous solution and its fertilizing effect on plants, which in turn produce organic acids in soils \citep{brady1991effect,brady1994direct}. In the remainder of this study, we ignore the impacts of life on weathering and focus on the ability of abiotic planets to achieve stable, temperate climates, since the impact of organisms on weathering may vary immensely from planet to planet, depending on specifics of metabolism and biogeochemical pathways. Increasing the CO$_2$ partial pressure of the atmosphere warms the planet and reduces the pH of rainwater, both of which should accelerate silicate dissolution under conditions where the kinetics of dissolution is relevant. This increases the delivery of Ca$^{2+}$ (or Mg$^{2+}$) cations to the ocean, which accelerates consumption of CO$_2$ and decreases pCO$_2$ until the rate of CO$_2$ consumption by silicate weathering is again equal to the rate of CO$_2$ production by outgassing. The opposite takes place in the case of a CO$_2$ reduction, so silicate weathering can act as a negative feedback on changes to the climate.

The temperature-dependence of silicate dissolution rates is usually represented as a simplified Arrhenius law:
\begin{linenomath}
\begin{align}
    r &\propto \exp{(\frac{T-T_{ref}}{T_e})}
\end{align}
\end{linenomath}
where $r$ is the silicate dissolution rate, $T$ is the temperature at which dissolution is taking place, and $T_{ref}$ is a reference temperature. $T_e = \frac{T^2_{ref}R}{E_{act}}$ is the temperature change required to increase or decrease the dissolution rate by a factor of $e$, where $R$ is the ideal gas constant and $E_{act}$ is the activation energy of the dissolution reaction. Under kinetically limited conditions, higher temperatures accelerate the dissolution of silicate minerals.

The pH-sensitivity of silicate dissolution is more complicated, as the magnitude and sign of the dependence often depends on the pH itself \citep[e.g.][]{brantley2008kinetics}. Under alkaline conditions, reduced pH often inhibits dissolution rates; near neutral conditions, the effect of pH is weak or non-existent; and in acidic conditions, reduced pH often accelerates dissolution. Under neutral-to-acid conditions, the silicate dissolution rate varies with pH as \citep[e.g.][]{kump2000chemical}:
\begin{linenomath}
\begin{align}
    r \label{eqn:dissolution-H-dep}&\propto [\text{H}^+]^{n_H}\\
    \log_{10}(r) &\propto -n_H\times \text{pH}
\end{align}
\end{linenomath}
where [H$^+$] is the activity or concentration of H$^+$ ions and $n_H$ is the reaction order with respect to H$^+$ ion activity. 

Following \citet{berner1992weathering}, in the limit where there are no influences on the water's pH but CO$_2$ concentration, the concentration of H$^+$ ions is determined by the reaction:
\begin{linenomath}
\begin{align}
    \text{CO}_2 + \text{H}_2\text{O} &\leftrightarrow \text{H}^+ +\text{HCO}_3^{-} 
\end{align}
\end{linenomath}
with an equilibrium constant of:
\begin{linenomath}
\begin{align}
    K_{eq} &= \frac{[\text{H}^+][\text{HCO}_3^-]}{[\text{CO}_2]}\\
    &=\frac{[\text{H}^+]^2}{[\text{CO}_2]}
\end{align}
\end{linenomath}
which implies:
\begin{linenomath}
\begin{align}
    [\text{H}^+] &\propto [\text{CO}_2]^{0.5}\\
    &\propto \text{pCO}_2^{0.5}
\end{align}
\end{linenomath}
where pCO$_2$ is the partial pressure of gaseous CO$_2$ in equilibrium with the silicate/water system, the last step following from Henry's Law. Combined with equation \ref{eqn:dissolution-H-dep}, this produces an expression for the pCO$_2$-dependence of silicate dissolution, assuming a water/rock system in equilibrium with gaseous CO$_2$ and without any other influences on pH:
\begin{linenomath}
\begin{align}
    r &\propto \text{pCO}_2^{0.5\times n_H}\\
    &\propto \text{pCO}_2^\beta
\end{align}
\end{linenomath}
where $\beta$ is simply 0.5$\times n_H$. It is important to recognize that the apparently direct dependence of dissolution rate on CO$_2$ is actually mediated by the impact of CO$_2$ on pH, and does not reflect a response to the abundance of CO$_2$ as a reactant, implying that other substances that influence pH in weathering systems could have just as much impact on weathering rates.

Combining the pCO$_2$-dependence and temperature-dependence presented above, the kinetically-limited dissolution rate of silicate minerals can be represented as:
\begin{linenomath}
\begin{align}\label{eqn:diss_rate}
    \frac{r}{r_{ref}} = \exp{(\frac{T-T_{ref}}{T_e})}(\frac{\text{pCO}_2}{\text{pCO}_{2,ref}})^\beta
\end{align}
\end{linenomath}
where the ``$ref$'' subscript refers to a reference value for a given variable. The values of $T_e$ and $\beta$, which determine the sensitivity of the reaction rate to changes in pCO$_2$ and temperature, vary considerably for different silicate minerals. We choose the default values listed in Table \ref{tab:values} based on results from laboratory silicate dissolution experiments like those cited after equation \ref{eqn:wollastonite}. 
\begin{table*}[htb!]
\centering
{
\begin{tabular}
{
l|c|c|r
}

\textbf{Parameter} & \textbf{Units} & \textbf{Definition} & \textbf{Default Value} \\
\hline
$\gamma$&$\textbf{--}$ & Land fraction & 0.3\\
\hline
$a_g$ & $\textbf{--}$ & Surface albedo & 0.2\\
\hline
$a$ & $\textbf{--}$ & Planetary albedo & 0.3\\
\hline
$R_{planet}$ & meters (m) & Planetary radius & 6.37$\times10^6$\\
\hline
$\gamma_{ref}$& $\textbf{--}$ & Modern land  &0.3\\
&&fraction&\\
\hline
$T_{ref}$& Kelvin (K) & Modern global-& 288\\
&&avg. temperature &\\
\hline
pCO$_{2,ref}$& bar & Pre-industrial &280$\times10^{-6}$\\
&&CO$_2$ partial pressure&\\
\hline
$p_{ref}$ & m yr$^{-1} $& Modern global- & 0.99\\
&&avg. precipitation&\citep{Xie-Arkin-1997:global}\\
\hline
$q_{ref}$& m yr$^{-1}$ & Modern global-&0.20\\
&&avg. runoff&\citep{oki2001global}\\
\hline
$\Gamma$ & $\textbf{--}$ & Fraction of precip.& $q_{ref}/p_{ref}= $0.2\\
&&that becomes runoff&\\
\hline
$\epsilon$& 1/K &Fractional change in& 0.03\\
&&precip. per K change in temp.&\\
\hline
$V_{ref}$ &mol yr$^{-1}$&Modern global& 7.5$\times 10^{12}$\\
&&CO$_2$ outgassing &\citep{gerlach2011volcanic,haqq2016limit}\\
\hline
$v$ & mol m$^{-2}$ yr$^{-1}$&Modern CO$_2$ outgassing& $V_{ref}/{4\pi R_{planet}^2} = 0.0147$\\
&&per m$^2$ planetary area&\\
\hline
$w_{w,ref}$ & mol m$^{-2}$ yr$^{-1}$&Modern weathering& $V_{ref}/{4\pi R_{planet}^2} = 0.0147$\\
&&per m$^2$ planetary area&\\
\hline
$\Lambda$&variable&Thermodynamic coefficient&1.4$\times10^{-3}$\\
&&for $C_{eq}$&\\
\hline
$n$&\textbf{--}&Thermodynamic pCO$_2$&0.316\\
&&dependence&\\
\hline
$\alpha$* & $\textbf{--}$ & $L\phi\rho_{sf}AX_r\mu$& 3.39$\times 10^5$\\
&&(see Section \ref{subsec:mac} and below)&\\
\hline
$L$*&m&Flow path length&1\\
\hline
$\phi$*&\textbf{--}&Porosity &0.1\\
\hline
$\rho_{sf}$*&kg m$^{-3}$&Mineral mass to&12728\\
&&fluid volume ratio&\\
\hline
$A$*&m$^2$ kg$^{-1}$&Specific surface&100\\
&&area&\\
\hline
$X_r$*&\textbf{--}&Reactive mineral conc.&0.36\\
&&in fresh rock&\\
\hline
$\mu$*&\textbf{--}&Scaling constant&e$^2$\\
\hline
$t_s$* & years & Soil age & $10^5$\\
\hline
$m$*&kg mol$^{-1}$&Mineral molar mass&0.27\\
\hline
$k_{eff,ref}$*&mol m$^{-2}$ yr$^{-1}$&Reference rate constant&8.7$\times10^{-6}$\\
\hline
$\beta$ & $\textbf{--}$ & Kinetic weathering & 0.2\\
& &pCO$_2$ dependence &\citep[e.g.][]{rimstidt2012systematic}\\
\hline
$T_e$ & Kelvin & Kinetic weathering & 11.1\\
& &temperature dependence  &\citep{Berner:1994p3295}\\
\hline
\end{tabular}
}
\caption{\textbf{List of parameters used in this study.} This table lists parameters used in our calculations, their units, their definitions, and the default values they take. A single asterisk (*) means the default parameter value was drawn from Table S1 of the supplement to \citet{maher2014hydrologic}. For default parameters drawn from other sources, the citation is given in the ``$\textbf{Value(s)}$'' column.
\label{tab:values}
}
\end{table*}
WHAK assumes that the global weathering rate is kinetically-limited, which would imply that increasing temperature or pCO$_2$ should always increase the flux of cations delivered to the ocean at a given runoff rate. This produces an equation of the form: 
\begin{linenomath}
\begin{align}\label{eqn:WHAK}
    \frac{W}{W_{ref}} &= \frac{Q}{Q_{ref}}\exp{(\frac{T-T_{ref}}{T_e})}(\frac{\text{pCO}_2}{\text{pCO}_{2,ref}})^\beta
\end{align}
\end{linenomath}
where $W$ is the global silicate weathering rate, $Q$ is the global runoff, and the $ref$ subscript refers to a reference value for a given variable. This formulation of global weathering adds another temperature dependence on top of the kinetic dependence, since runoff increases with precipitation, which increases with global temperature \citep[e.g.][]{held2006robust,o2008hydrological}. Some formulations of the weathering feedback \citep[e.g.][]{Berner:1994p3295,Pierrehumbert:2010-book,abbot12-weathering} use $(\frac{Q}{Q_{ref}})^a$ where $a = 0.65$ based on \citet{DUNNE:1978p3297,Peters84}, but \citet{abbot12-weathering} demonstrated that weathering behavior in the WHAK model is not sensitive to the choice of $a$ for values between 0 and 2, so we will use $a=1$ for simplicity. 

Rocky exoplanet climate and habitability studies that include silicate weathering invariably use the WHAK model \citep[e.g.][]{Kite:2011,abbot12-weathering,edson2012carbonate,kadoya2012climate,watanabe2012climate,kadoya2014conditions,Menou2015,foley2015role,abbot-2016,batalha2016climate,haqq2016limit,paradise2017,ramirez2018more,rushby2018long,kadoya2019outer,paradisehabitable,checlair2019no}. However, there are orders-of-magnitude discrepancies between the weathering rates predicted by laboratory silicate dissolution experiments and the weathering rates observed in field studies of silicate weathering \citep{velbel1993constancy,malmstrom2000resolving, white2003effect,maher2006mineral}. Further, despite a fairly stable climate over the past 3-4 Ga given the long-term trend in solar forcing, Earth has varied between ``hothouse'', ``icehouse'', and even snowball states, which could be explained partially by variations in the strength of the negative feedback on Earth's climate \citep{kump1997global,west2005tectonic,maher2014hydrologic}. 
\subsection{The MAC Model}\label{subsec:mac}
The ``solute transport model'' presented in \citet{maher2014hydrologic} (MAC) attempts to provide a mechanistic explanation for variations in the strength of the silicate weathering feedback on Earth through time and in different river catchments. The solute transport model is the first model of the silicate weathering feedback to explicitly include a ``thermodynamic limit'', $C_{eq}$, on cation concentration in runoff, which emerges due to the control on silicate mineral dissolution by precipitation of secondary minerals like clays \citep{alekseyev1997change,maher2006mineral,maher2009role,winnick2018relationships}. When silicate dissolution and clay precipitation reach thermodynamic equilibrium, the maximum solute concentration of silicate dissolution products is achieved, maximizing weathering flux for a given runoff. In the MAC framework, for river catchments draining silicate mineral assemblages, the flux of cations delivered to the ocean (and therefore the CO$_2$ sequestration potential) depends on the ratio between the mean fluid travel time through a reactive assemblage ($t_f\approx L\phi/q$) and the time required for the system to reach $C_{eq}$, its maximum concentration ($t_{eq}\approx C_{eq}/r_{n}$); $L$ is the reactive flow path length (which may vary with soil thickness \citep{ferrier2008effects,gabet2009theoretical}, although this depends on the setting where the bulk of weathering takes place \citep{west2012thickness,carretier2018colluvial}; see also Section S.1 in the supplement to \citet{maher2014hydrologic}), $\phi$ is the effective porosity of the minerals the water is flowing through, $r_n$ is the mineral reaction rate, and $q$ is the runoff. $r_n = \rho_{sf}k_{eff}AX_rf_w$, where $\rho_{sf}$ is the ratio of the mass of solid mineral to volume of fluid, $k_{eff}$ is the dissolution rate constant, $A$ is the specific surface area for the minerals in the assemblage, $X_r$ is the fraction of reactive minerals in fresh, unweathered soil/rock, and $f_w$ is the fraction of fresh, unweathered soil/rock in the assemblage. $f_w=\frac{1}{1+mk_{eff}At_s}$, where $m$ is the molar mass of the minerals and $t_s$ is the age of the soil. Note that the fraction of unweathered minerals decreases with soil age, and that a higher weathering constant ($k_{eff}$) will lead to a smaller fraction at a given soil age, depleting the soil/rock of reactive minerals. The ratio of mean fluid travel time through an assemblage and mean time to reach thermodynamic equilibrium ($t_f/t_e$) is known as  the Damk\"{o}hler number ($Da$) \citep{boucher1963dimensionless}; MAC factor out runoff to introduce the ``Damk\"{o}hler coefficient'' ($D_w$):
\begin{linenomath}
\begin{align}
    D_w &=\label{eqn:Dw} \frac{L\phi\rho_{sf}k_{eff}AX_r}{C_{eq}(1+mk_{eff}t_s)}
    \end{align}
\end{linenomath}
The weathering rate constant $k_{eff}$ can be described by an equation with the same functional form as our equation \ref{eqn:diss_rate}, e.g. $\frac{k_{eff}}{k_{eff,ref}} = \exp{(\frac{T-T_{ref}}{T_e})}(\frac{\text{pCO}_2}{\text{pCO}_{2,ref}})^\beta$. $D_w$ is central to the MAC formulation of weathering; the higher the value of $D_w$ for a mineral assemblage being drained by a given flux of runoff, the higher the concentration of cations in the runoff, up to the thermodynamic limit on concentration, $C_{eq}$. Further, a higher $D_w$ value means that an increase in runoff leads to less dilution of solute, allowing for larger changes to weathering rates in response to climate perturbations.

\citet{winnick2018relationships} showed that $C_{eq}$ should be dependent on pCO$_2$. As discussed in Section \ref{subsec:whak}, pCO$_2$ can be an important control on the pH of water at Earth's surface; this provides a mechanism for changes in pCO$_2$ to change the equilibrium concentrations of reactants and products in mineral dissolution/precipitation reactions. Following 
\citet{winnick2018relationships}, this can be illustrated with a theoretical silicate dissolution / clay precipitation reaction:
\begin{linenomath}
\begin{align}
    dD + 2\text{CO}_2 (g) + \text{H}_2\text{O} \leftrightarrow sS + 2\text{HCO}_3^- + aA + bB + c\text{SiO}_2 (aq)
\end{align}
\end{linenomath}
where $D$ is the dissolving mineral (e.g. plagioclase feldspar), $A$ and $B$ respectively are the divalent and monovalent weathering products (e.g. Ca$^{2+}$ and Na$^{+}$), $S$ is the precipitating secondary mineral (e.g. halloysite), and lower-case letters are stoichiometric coefficients. We can then write the equilibrium constant for the reaction as:
\begin{linenomath}
\begin{align}
K &= \frac{[\text{HCO}_3^{-}]^2[A]^a[B]^b[C]^c}{(\text{pCO}_2)^2}\\
&= \frac{z[A]^{2+a+b+c}}{(\text{pCO}_2)^2}
\end{align}
\end{linenomath}
where $z = (\frac{2}{a})^2(\frac{b}{a})^b(\frac{c}{a})^c$ captures the relative stoichiometric coefficients and the second step assumes that no other reactions influence the concentrations of the reactants. Now we can write $C_{eq}$ in terms of pCO$_2$, $z$, $K$, and the stoichiometric coefficients:
\begin{linenomath}
\begin{align}
    [A] = C_{eq} &= (\frac{K}{z})^{\frac{1}{2+a+b+c}}(\text{pCO}_2)^{\frac{2}{2+a+b+c}}\label{eqn:n_calc}\\
    &=\Lambda(\text{pCO}_2)^n\label{eqn:Ceq}
\end{align}
\end{linenomath}
where $[A]$, the divalent cation concentration, is chosen for $C_{eq}$ since the delivery of Ca$^{2+}$ to the oceans by silicate weathering is thought to be the dominant sink for CO$_2$ on geologic timescales through CaCO$_3$ formation and burial \citep{france1997organic,sun2016diffusive}. The temperature-dependence of $C_{eq}$ is weak (see below and \citet{winnick2018relationships}), so we will ignore it in this study. This is a powerful formulation because it allows the pCO$_2$ feedback strength to be derived directly from equilibrium chemistry. However, it is also a simplification, since the concentrations of reactants in mineral dissolution and precipitation reactions will be influenced by other, coupled reactions and particularly by the presence of other sources of acidity and alkalinity. We also note that we have only modeled ``open-system'' weathering in this study, meaning we have assumed that the weathering system is always in equilibrium with the ambient air and is thus continually recharged with CO$_2$ as weathering takes place. In the limit of ``closed-system'' weathering, weathering takes place in a system that is not recharged with CO$_2$. These two formulations of weathering display different behavior at low pCO$_2$, with $C_{eq}$ varying linearly with CO$_2$ in the closed system case. However, \citet{winnick2018relationships} demonstrate that their behavior generally converges at pCO$_2$ $\leq10^{-1}$ bar, and a slightly different functional form of the relationship between pCO$_2$ and $C_{eq}$ at low pCO$_2$ would not impact the qualitative results of this study. 

Still following \citet{winnick2018relationships}, we will use the dissolution of plagioclase feldspar (An20) and precipitation of halloysite as the example reaction to derive default values for the coefficient and exponent in equation \ref{eqn:Ceq}: 
\begin{linenomath}
\begin{align}
\begin{split}
    &1.66\text{Ca}_{0.2}\text{Na}_{0.8}\text{Al}_{1.2}\text{Si}_{2.8}\text{O}_8 + 2\text{CO}_2 (g) + 3\text{H}_2\text{O}\\
    &\leftrightarrow \text{Al}_2\text{Si}_2\text{O}_5\text{(OH)}_4 + 0.33\text{Ca}^{2+}+1.33\text{Na}^+ + 2\text{HCO}_3^-\\
    &+2.66\text{SiO}_2 (aq)
\end{split}
\end{align}
\end{linenomath}
which gives $n=0.316$ and an equilibrium constant of $K=$10$^{-13.3}$ at 288 K, implying $\Lambda=0.0014$. As noted above, $K$ is weakly temperature-dependent, producing variations in $\Lambda$ with temperature: for the above reaction, $\Lambda(273 \text{ K})=0.0011$ and $\Lambda(373\text{ K})=0.0033$. This effect provides a weak negative feedback on climate by increasing $C_{eq}$ with temperature, but simulations we carried out that included this temperature-dependence (not shown) differed insignificantly from those that excluded it, so for simplicity we did not include this effect in the simulations shown in this study. It is also important to note that different mineral assemblages can have a large range of values for $n$, and $\Lambda$ can span orders of magnitude for different lithologies \citep{winnick2018relationships}, so we vary these parameters widely in our calculations. 

\begin{figure*}[htb!]
    \centering
    \makebox[\textwidth][c]{\includegraphics[width=400pt,keepaspectratio]{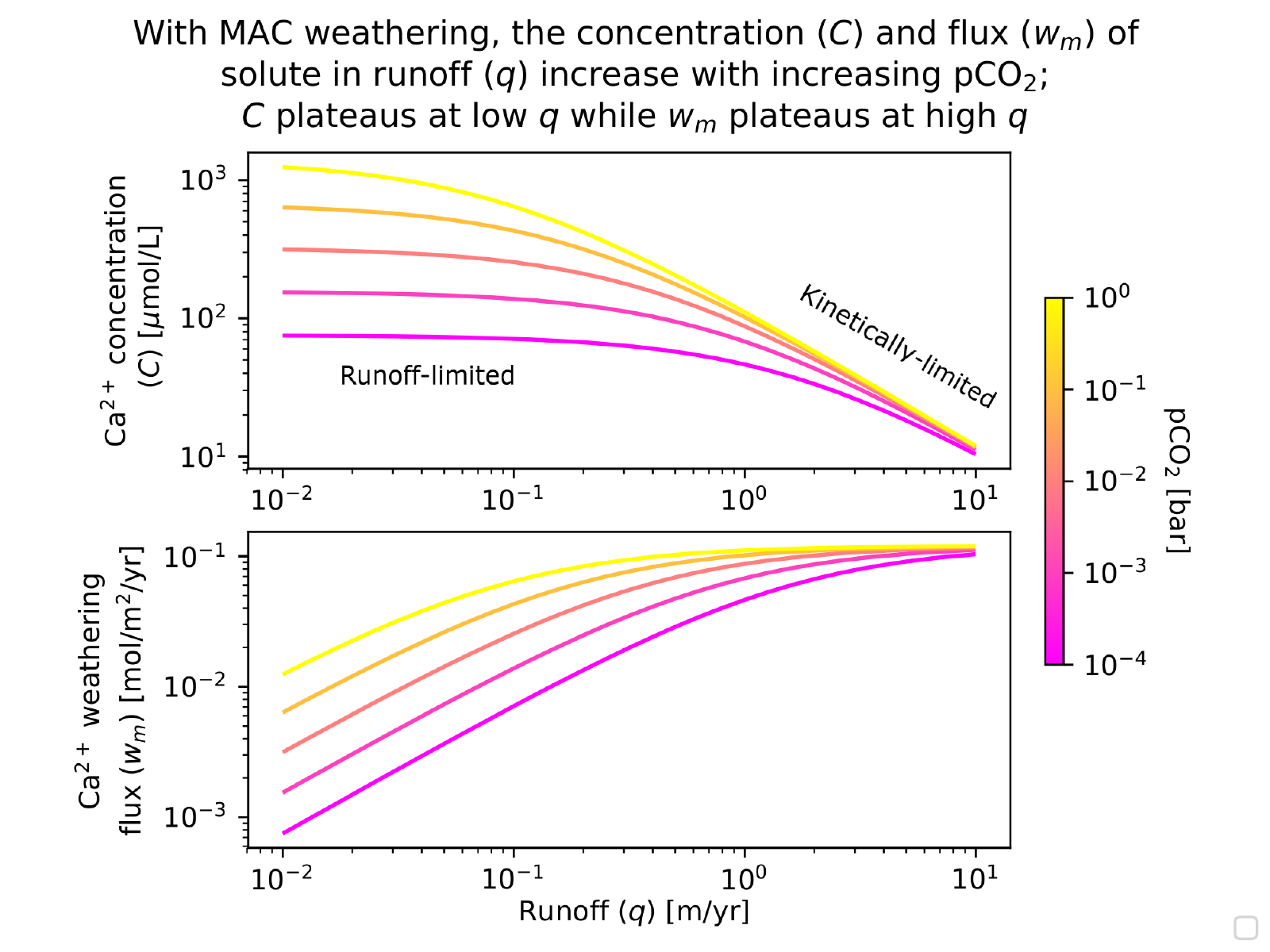}}
    \caption{Solute concentrations and weathering fluxes as functions of runoff and pCO$_2$ with our default parameter choices (see Table \ref{tab:values}). The top plot shows Ca$^{2+}$ concentrations (equation \ref{eqn:mac_conc}) and the bottom plot shows Ca$^{2+}$ fluxes to the ocean (equation \ref{eqn:OG_mac}). The pinkest curves have a $C_{eq}$ calculated with pCO$_2=10^{-4}$ bar (see equation \ref{eqn:Ceq}), and the color of the curves shifts from pink to yellow as pCO$_2$ increases from 10$^{-4}$ to 10$^{-3}$ to 10$^{-2}$ to 10$^{-1}$ to 10$^{0}$ bar CO$_2$. The regions with runoff limitation and kinetic limitation are labeled as such.}
    \label{fig:concentration_flux}
\end{figure*}

Using equations \ref{eqn:Dw} and \ref{eqn:Ceq} for $D_w$ and $C_{eq}$ and the weathering and solute concentration equations given in MAC, we can derive a version of MAC's weathering equation that includes the impact of temperature and pCO$_2$ on dissolution kinetics and the impact of pCO$_2$ on thermodynamic equilibrium solute concentration (see the supplement to \citet{maher2014hydrologic} for a full derivation of equation \ref{eqn:mac_conc}): 
\begin{linenomath}
\begin{align}
    C &= \label{eqn:mac_conc} C_{eq}\frac{\mu D_w/q}{1+\mu D_w/q}\\
    w &= \label{eqn:OG_mac}qC_{eq}\frac{\mu D_w/q}{1+\mu D_w/q}\\
    &=\label{eqn:mac_weathering}\frac{\alpha}{[k_{eff,ref} \exp{(\frac{T-T_{ref}}{T_e})}(\frac{\text{pCO}_2}{\text{pCO}_{2,ref}})^\beta]^{-1}+mAt_s+\alpha[(qC_{eq})]^{-1}}
\end{align}
\end{linenomath}
where $C$ is the solute concentration in runoff, $w$ is weathering per unit surface area, $\alpha=L\phi\rho_{sf}AX_r\mu$ for compactness of notation, and $\mu=e^2$ (see supplement to MAC, where $\mu$ is instead called $\tau$). Through its thermodynamic impact on $C_{eq}$ (equation \ref{eqn:Ceq}), pCO$_2$ has a strong effect on the concentration of solute in runoff, and thus on the weathering flux and CO$_2$ sink at a given runoff. Figure \ref{fig:concentration_flux} shows the solute concentrations and weathering fluxes for the weathering assemblage described in the previous paragraph across a wide range of runoff and pCO$_2$.
In the limit where $\frac{1}{mAt_s+\alpha/(qC_{eq})} \gg k_{eff}$, equation \ref{eqn:mac_weathering} displays the ``kinetically-limited'' weathering behavior assumed in the WHAK model. This happens when $t_s$ is small and $q$ or $C_{eq}$ is large -- when relatively fresh, unweathered soil is being flushed out by large enough volumes of runoff to maintain very dilute solute concentrations. When $mAt_s$ is large enough relative to other terms not to neglect, $\alpha/(qC_{eq})\sim 0$, and $[k_{eff}]^{-1}\sim0$, e.g. when runoff and the weathering rate constant are both large but $t_s$ is non-negligible so the soil being flushed out by the runoff is already partially weathered, equation \ref{eqn:mac_weathering} limits to a constant value of $\frac{\alpha}{mAt_s}$, which goes to zero as $t_s$ grows very large. Weathering in this limit is ``supply-limited,'' since the weathering rate is dictated by the rate of supply of fresh material to the weathering zone, with a lower $t_s$ implying a higher rate of delivery and a less stringent supply limit. When $qC_{eq}\ll \alpha/(mAt_s)$ and $qC_{eq}\ll \alpha k_{eff}$, e.g. when runoff flushes the system out slowly enough relative to other factors that the solute can approach its thermodynamic limit on concentration $C_{eq}$, the weathering rate varies linearly with runoff. In this regime, the system displays ``runoff-limited'' (also referred to as ``chemostatic'') weathering, where the kinetics of silicate dissolution do not influence weathering rate, and weathering flux is controlled entirely by runoff at a given pCO$_2$. 

\textcolor{black}{For context, we will briefly discuss where Earth lies in terms of kinetic vs. runoff controls on global weathering flux, using results from studies that apply the MAC framework to interpret proxies and direct weathering measurements. By averaging the annual-mean silica concentrations and fluxes from a dataset of large rivers draining granitic lithologies \citep{gaillardet1999global}, \citet{maher2014hydrologic} estimate that Earth as a whole lies in the transition zone between fully runoff-limited and kinetically-limited behavior, with a high degree of variation between individual rivers (see their Figures 2 and 3). \citet{von2015stable} combine general circulation model simulations and a beryllium weathering proxy to derive a similar result for Earth's weathering regime since the Last Glacial Maximum. Additionally, using non-averaged solute concentration vs. runoff measurements paired with individual estimates of $C_{eq}$ for rivers draining both granitic and basaltic lithologies, \citet{ibarra2016differential} showed that the median concentration-runoff pairs for many individual rivers also lie in the transition regime, though they are on average closer to the runoff limit than the global average estimate by \citet{maher2014hydrologic}. Finally, we note that most of the cases examined in this paper lie in the runoff-limited regime or near the transition to kinetic limitation, like Earth.}

\section{Model Description $\&$ Methods}\label{sec:methods}
Here we describe the models we used to evaluate the consequences of thermodynamic and energetic limits on silicate weathering for the climate stability of rocky land- and ocean-bearing planets in the habitable zone. We coupled a zero-dimensional energy balance model to a zero-dimensional CO$_2$ balance model to simulate the first-order behavior of planets under a variety of conditions. This clearly requires extreme simplification of planetary processes, but this is justified given the lack of observational information about habitable zone worlds at present. 

The energy balance model equates the global average of absorbed instellation with outgoing longwave radiation ($OLR$):
\begin{linenomath}
\begin{align}
    S_{avg} &= \frac{1}{4}(1-a)S\\
    &= OLR
\end{align}
\end{linenomath}
where $S_{avg}$ is the globally-averaged absorbed instellation, $S$ is the top-of-atmosphere instellation at the substellar point, $a$ is the planetary albedo, and $1/4$ is a geometric factor accounting for the fact that a planet's surface area is 4$\times$ the area of its cross-section. 

By default, our simulations assume a constant planetary albedo $a=0.3$, approximately corresponding to Earth's present-day value \citep{stephens2015albedo}. We note that a major uncertainty in our study is the exclusion of interactive cloud effects, which strongly impact planetary albedo and energy balance (see Section \ref{subsubsec:limitations} for discussion of this point). We also include a set of simulations with planetary albedo that is a function of surface albedo as well as atmospheric albedo due to Rayleigh scattering by pCO$_2$. This allows us to approximate the impact of atmospheric scattering on the climates of planets orbiting G2-stars like the Sun; the planetary albedos of worlds orbiting K- and M-dwarfs are less affected by Rayleigh scattering, since those classes of stars emit light at longer wavelengths where atmospheres are more strongly absorbing and display weaker scattering \citep[e.g.][]{Kopparapu:2013}. The equations for calculating Rayleigh scattering planetary albedo under a two-stream Eddington approximation are \citep{Pierrehumbert:2010-book}:
\begin{linenomath}
\begin{align}
    \tau_{ray} &=0.19513\frac{\text{pCO}_2}{1\text{ bar CO}_2}\\
    a_a &= \frac{(0.5-0.75\cos{\zeta})(1-\exp(-\tau_{ray}/\cos{\zeta}))+0.75\tau_{ray}}{1+0.75\tau_{ray}}\\
    a_a' &= \frac{0.75\tau_{ray}}{1+0.75\tau_{ray}}\\
    a &= 1 - \frac{(1-a_g)(1-a_a)}{(1-a_g)a_a'+(1-a_a')}\label{eqn:planetary_albedo}
\end{align}
\end{linenomath}
where $\tau_{ray}$ is the Rayleigh scattering optical depth as a function of pCO$_2$ on a planet with surface gravity equal to Earth's at wavelength 0.5 micron, which is near the peak of the solar spectrum (drawn from Table 5.2 in \citet{Pierrehumbert:2010-book}), $a_a$ is the atmospheric albedo due to Rayleigh scattering of incoming solar radiation, $\cos{\zeta}$ is the cosine of the zenith angle of the star (taken to be 2/3 in this study, based on \citet{cronin2014choice}), $a_a'$ is the atmospheric albedo for upward-directed diffuse radiation from the surface, $a_g$ is the surface albedo, and the equation for $a$ combines $a_a$, $a_a'$, and $a_g$ to get total planetary albedo. In the subset of experiments that include this representation of Rayleigh scattering, we vary $a_g$ from 0.1 to 0.3, covering a range of plausible surface albedos that Earth may have displayed throughout its history as cloud and continental coverage evolved \citep{Rosing:2010p1227}.

$OLR$ is calculated using a polynomial fit presented in \citet{kadoya2019outer} that approximates the output of the 1-D radiative-convective model used in \citet{Kopparapu:2013} with 1 bar N$_2$ and saturated H$_2$O for temperatures between 150 K and 350 K and pCO$_2$ between 10$^{-5}$ bar and 10 bar:
\begin{linenomath}
\begin{align}
    OLR(T,\text{pCO}_2)\label{eqn:olr} = &I_0 + \textbf{TB$\Upsilon$}^t\\
    \textbf{T} = (1\;\xi\;\xi^2\;\xi^3&\;\xi^4\;\xi^5\;\xi^6)\\
    \textbf{$\Upsilon$} = (1\;\upsilon\;\upsilon^2&\;\upsilon^3\;\upsilon^4)
\end{align}
\end{linenomath}
where $T$ is the surface temperature in K, pCO$_2$ is the partial pressure of CO$_2$ in bars, $I_0=-3.1$ W m$^{-2}$, and $\xi=0.01\times(T-250)$. For pCO$_2<1$ bar:
\begin{linenomath}
\begin{align}
    &\;\;\;\;\;\;\;\;\;\;\;\;\;\;\;\;\;\;\;\;\;\;\;\;\upsilon = 0.2\times \log_{10}(\text{pCO}_2)\\
    \textbf{B}&=
    \begin{bmatrix}
        87.8373 &311.289& 504.408& 422.929& 134.611\\
        54.9102& 677.741& 1440.63& 1467.04& 543.371\\
        24.7875& 31.3614& 364.617& 747.352& 395.401\\
        75.8917& 816.426& 1565.03& 1453.73& 476.475\\
        43.0076& 339.957& 996.723& 1361.41& 612.967\\
        31.4994& 261.362& 395.106& 261.600& 36.6589\\
        28.8846& 174.942& 378.436& 445.878& 178.948
    \end{bmatrix}
\end{align}
\end{linenomath}

For 10 bar $>$ pCO$_2>1$ bar:
\begin{linenomath}
\begin{align}
    &\;\;\;\;\;\;\;\;\;\;\;\;\;\;\;\;\;\;\;\;\;\;\;\;\;\;\;\;\upsilon = \log_{10}(\text{pCO}_2)\\
    \textbf{B}&=
    \begin{bmatrix}
        87.8373& 52.1056& 35.2800& 1.64935& 3.42858\\
        54.9102& 49.6404& 93.8576& 130.671& 41.1725\\
        24.7875& 94.7348& 252.996& 171.685& 34.7665\\
        75.8917& 180.679& 385.989& 344.020& 101.455\\
        43.0076& 327.589& 523.212& 351.086& 81.0478\\
        31.4994& 235.321& 462.453& 346.483& 90.0657\\
        28.8846& 284.233& 469.600& 311.854& 72.4874
    \end{bmatrix}
\end{align}
\end{linenomath}
This simple parameterization of OLR allows us to perform rapid simulations across a wide range of parameters, with a maximum absolute error of 3.3 W m$^{-2}$ and an average absolute error of 0.6 W m$^{-2}$ in the range of temperatures and pCO$_2$ mentioned above \citep{kadoya2019outer}. The main drawback in using this parameterization is that we cannot evaluate climates at pCO$_2$ above 10 bar or with varying N$_2$ abundance. 

The CO$_2$ balance model sets CO$_2$ outgassing from volcanism equal to CO$_2$ consumption from weathering:
\begin{linenomath}
\begin{align}
    v &= w \label{eqn:volc}
\end{align}
\end{linenomath}
where $v = V/(4\pi R_{planet}^2)$ is the total volcanic outgassing of CO$_2$ ($V$) divided by the planetary surface area, $R_{planet}$ is the radius of the planet, and $w$ is the CO$_2$ consumption by weathering per unit planetary surface area. To compare the behavior of planets with the WHAK and MAC weathering models, we run simulations with each formulation:
\begin{widetext}
\begin{linenomath}
\begin{align}
w_w &= \label{eqn:whak_weather} w_{w,ref}\frac{\gamma p}{\gamma_{ref}p_{ref}}\exp{(\frac{T-T_{ref}}{T_e})}(\frac{\text{pCO}_2}{\text{pCO}_{2,ref}})^\beta\\
w_m \label{eqn:mac_weather}&=\gamma\frac{\alpha }{[k_{eff,ref} \exp{(\frac{T-T_{ref}}{T_e})}(\frac{\text{pCO}_2}{\text{pCO}_{2,ref}})^\beta]^{-1}+mAt_s+\alpha[q\Lambda(\text{pCO}_2)^n]^{-1}}
\end{align}
\end{linenomath}
\end{widetext}
where $w_w$ is the weathering per unit planetary area using the WHAK model; $\gamma$ is the global land fraction, which is included to account for the fact that continental silicate weathering only happens on continents; $p$ is the global average precipitation per unit area; and $w_m$ is the weathering per unit planetary area using the MAC model. Crucially, the impact of the kinetic dissolution rate and its dependence on temperature and pCO$_2$ is greatly weakened in equation \ref{eqn:mac_weather}. This is because increased dissolution of silicates also depletes soils and rocks of their reactive components more completely for a given soil age, reducing the reactivity of the assemblage and counteracting the increase in dissolution rate. However, pCO$_2$ still has a powerful impact on silicate weathering through its control of the thermodynamic equilibrium concentration of solute in runoff (equation \ref{eqn:Ceq}). 

We take the runoff $q$ to be a linear function of precipitation $p$, which is itself taken to be a linear function of temperature $T$ \citep[e.g.][]{held2006robust, o2008hydrological}:
\begin{linenomath}
\begin{align}
    q &= \Gamma \times p\label{eqn:runoff}\\
    p &= p_{ref}(1+\epsilon(T-T_{ref}))\label{eqn:precip}
\end{align}
\end{linenomath}
where  $\Gamma$ is the proportionality constant between precipitation and runoff and $\epsilon$ is the fractional change in global precipitation per Kelvin deviation from a reference surface temperature. 

We also include the upper limit on precipitation noted by \citet{Pierrehumbert-2002:hydrologic,o2008hydrological}. In steady-state, global precipitation is equal to global evaporation. Global evaporation is ultimately driven by instellation, which provides the energy required for water to transition from liquid to gas (the latent heat of vaporization). If there is not enough instellation to make up for the cooling caused by a given amount of evaporation, then a planet cannot sustain that level of precipitation. This implies a maximum global precipitation given by:
\begin{linenomath}
\begin{align}
    p_{lim} &= \frac{S_{avg}}{L(T)}\label{eqn:plim}\\
    L(T) &= 1.918\times10^9(\frac{T}{T-33.91})^2
\end{align}
\end{linenomath}
where $p_{lim}$ is the maximum precipitation sustainable on a planet at a given instellation and $L(T)$ is the latent heat of vaporization for water in J m$^{-3}$, converted from $L$ in J kg$^{-1}$ given by the Henderson-Sellers equation \citep{henderson1984new} with multiplication by water's density 1000 kg m$^{-3}$. 

More speculatively, we propose that the actual value of $p_{lim}$ at a given instellation may be lower than that given by equation \ref{eqn:plim} for planets with land, since stellar energy absorbed by continents can only recycle water that ultimately evaporated from the ocean (note that the validity of equation \ref{eqn:plim} was only demonstrated in the case of a fully ocean-covered planet by \citet{o2008hydrological}). Although the influence of a given amount of land area on the energetic precipitation limit should depend on the latitude of the land mass (e.g. a continent will intercept many more photons if it straddles the equator or substellar point than if it sits on a pole or at the terminator) and the efficiency of the atmosphere at moving energy from the land surface to the ocean surface, we suggest an approximate, first-order scaling of $p_{lim}$ with global ocean fraction (1-$\gamma$), such that equation \ref{eqn:plim} represents $p_{lim}$ in the 100$\%$ ocean coverage limit:
\begin{linenomath}
\begin{align}
    p_{lim,land} \label{eqn:plimland}&= (1-\gamma)\frac{S_{avg}}{L(T)}
\end{align}
\end{linenomath}
where $p_{lim,land}$ is the energetic limit on precipitation for planets with a non-zero surface land fraction. Since we have not yet demonstrated the validity of equation \ref{eqn:plimland} explicitly with GCM simulations (a subject for future work), we use equation \ref{eqn:plim} for the energetic limit in most calculations, but we also carry out a set of experiments with equation \ref{eqn:plimland} to examine the impact of the proposed scaling.

We search for equilibrium pCO$_2$ and $T$ for planets at $S$ from 1400 W m$^{-2}$ to 473 W m$^{-2}$, the outer edge of the classical habitable zone for an Earth-sized planet orbiting a G-star according to \citet{Kopparapu:2013}, defined by the ``maximum greenhouse'' limit where Rayleigh scattering from CO$_2$ accumulation begins to outweigh CO$_2$'s warming effect. Note that we do \textit{not} have a representation of Rayleigh scattering in our default model, so we take the value of the outer edge given in \citet{Kopparapu:2013}, though we also explore the impact of a Rayleigh scattering parameterization in a separate set of simulations. To find equilibrium values, we solve for pCO$_2$ and $T$ where $v=w$ and $S_{avg}=OLR$ simultaneously. Using this procedure to find steady-state climates, we compare the behavior of planets with weathering described by the MAC formulation against that of planets under the WHAK formulation and vary parameters across a range of values to examine the sensitivity of our models. In particular, we examine the pCO$_2$ and temperatures of our simulations as a function of instellation, and we calculate the ``effective outer edge of the habitable zone'' as the instellation at which the temperature of a simulation with a given set of parameters drops below freezing (while acknowledging that planets may be able to reach global average temperatures somewhat below freezing before entering a ``hard snowball state'' \citep[e.g.][]{Abbot-et-al-2011:Jormungand,yang2012}). In the next section we present the results of these calculations.
\section{Results}\label{sec:results}
We begin by comparing the simulations with the WHAK formulation against the simulations with the MAC formulation. The different weathering models lead to qualitative differences in weathering behavior, planetary climate, and the effective outer edge of the HZ. Then, we examine the impact of varying the parameters that represent hydrology, weathering thermodynamics, surface properties, and albedo in our model of the MAC formulation of weathering. The effective outer edge of the habitable zone is quite sensitive to all of these sets of parameters. This makes it difficult to predict what fraction of rocky, ocean-bearing planets in the HZ we should expect to find in a temperate state instead of a snowball or a moist greenhouse, even if we ignore the strong possibility that other greenhouse gases or CO$_2$ cycling mechanisms will be present on putatively Earth-like planets. Finally, we examine the impact of the energetic limit on precipitation set by planetary instellation (see equations \ref{eqn:plim} and \ref{eqn:plimland}), which we find fundamentally changes the functioning of the silicate weathering feedback under some circumstances by decoupling planetary temperature and global runoff. 
\subsection{WHAK vs. MAC}\label{subsec:whakvmac}

\subsubsection{Sensitivity to land fraction and CO$_2$ outgassing rate}
\begin{figure*}[htb!]
    \centering
    \makebox[\textwidth][c]{\includegraphics[width=400pt,keepaspectratio]{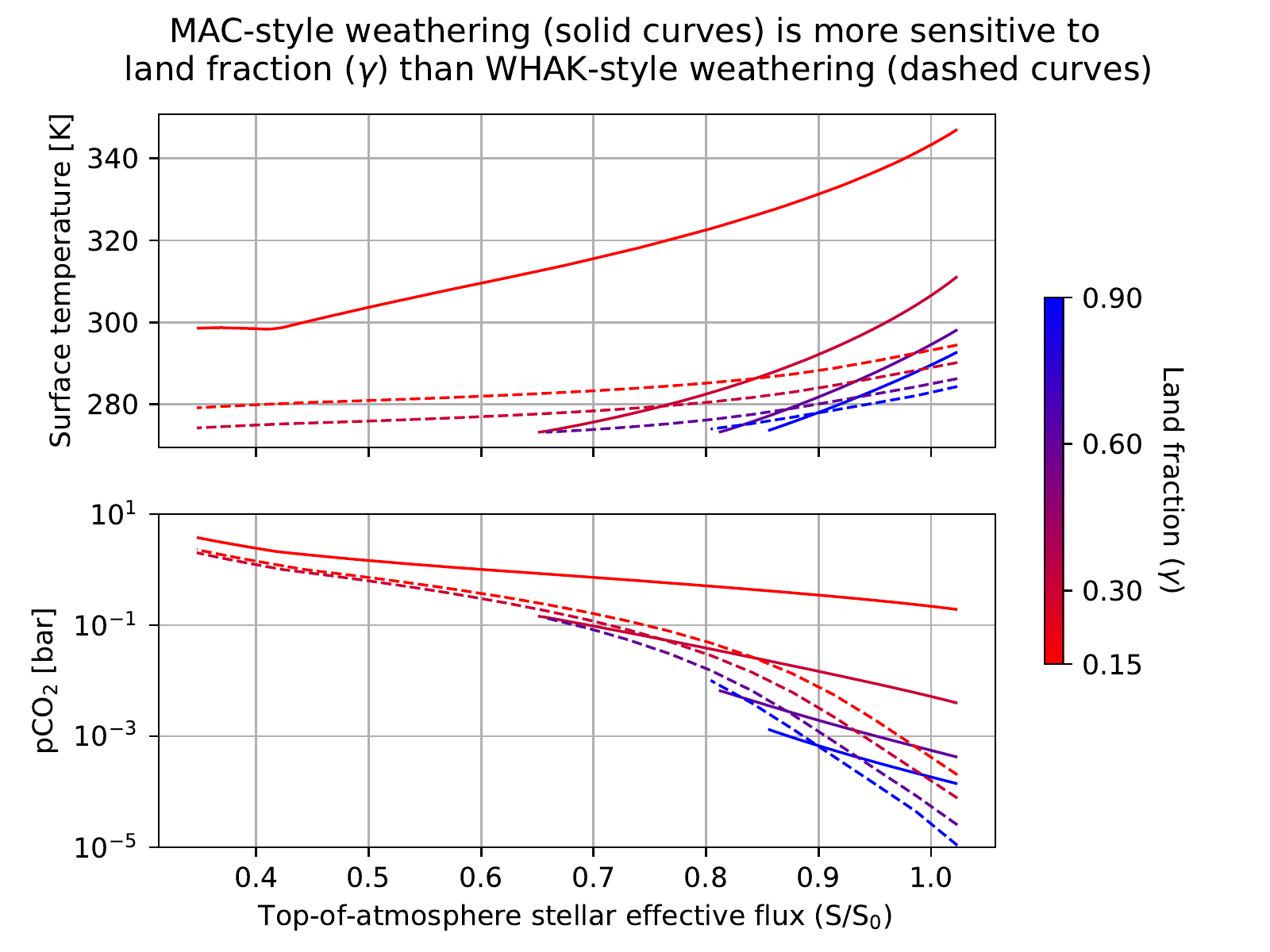}}
    \caption{A comparison of instellation vs. temperature and pCO$_2$ for WHAK and MAC simulations at varying land fraction. The top plot shows top-of-atmosphere stellar effective flux ($S/S_0$, where $S_0=1368$ W m$^{-2}$) vs. temperature. The bottom plot shows TOA stellar effective flux vs. pCO$_2$. Solid curves use the MAC formulation of weathering, and dashed curves use the WHAK formulation. The reddest curves have land fraction ($\gamma$)=0.15. $\gamma$=0.3 curves are red-purple, $\gamma=0.6$ curves are blue-purple, and $\gamma=0.9$ curves are blue.}
    \label{fig:land_frac_whak_mac}
\end{figure*}

As demonstrated in \citet{abbot12-weathering, foley2015role}, weathering behavior with the WHAK model is fairly insensitive to planetary land fraction; in contrast, MAC planet climates are considerably more sensitive to land fraction, particularly at $\gamma < 0.3$ (see Fig. \ref{fig:land_frac_whak_mac} and the leftmost column in Fig. \ref{fig:whak_vs_mac}). The MAC simulation with $\gamma=0.15$ is 30-40 K warmer than the simulation with $\gamma=0.3$ at all instellations, and with $\gamma<0.15$, we failed to find stable solutions within the the temperature and pCO$_2$ range of the OLR parameterization we are using, though stable climates conceivably exist with pCO$_2>10$ bar and/or $T>350$ K. Also, other combinations of parameters would lead to different values in either direction for the minimum $\gamma$, so $\gamma=0.15$ is not a hard minimum on the land fraction a planet needs to stabilize its climate via silicate weathering. Interestingly, Fig. \ref{fig:land_frac_whak_mac} shows that the temperature of the $\gamma=0.15$ case stops decreasing with instellation when $S/S_0$ falls below $\sim0.4$ (see red curve in top panel of Fig. \ref{fig:land_frac_whak_mac}). This is because that set of simulations hits the energetic limit on precipitation at that instellation (see Section \ref{subsec:energetic_limit} for further analysis of this phenomenon). 
Since $v=w$, and $\gamma$ is a multiplicative factor on the front of the equations for $w_w$ and $w_m$ (equations \ref{eqn:whak_weather} and \ref{eqn:mac_weather}), a fractional change in $\gamma$ at constant $v$ has the same effect as multiplying $v$ by the reciprocal of that fraction while holding $\gamma$ constant. So, planets under the MAC formulation are also sensitive to increases in CO$_2$ outgassing rates, as reducing $\gamma$ by a factor of 2 from 0.3 to 0.15 is equivalent to multiplying volcanic outgassing by 2 while holding land fraction constant. 

The difference in land fraction and outgassing sensitivity between MAC and WHAK planets is due to the lack of weathering sensitivity to dissolution kinetics in the MAC planets. Under decreased land fraction (or increased outgassing), more weathering must happen per unit land area to balance outgassing. Without temperature-dependent changes in solute concentration via changes in dissolution kinetics, MAC planets are forced to compensate for changes in land fraction or outgassing through changes to precipitation rate (which responds to temperature) and maximum solute concentration $C_{eq}$ (which responds to pCO$_2$). On WHAK planets, the solute concentration can increase or decrease without limit in response to both temperature and pCO$_2$ due to changes in the kinetic silicate dissolution rate. This allows for greater changes in concentration to bolster changes in runoff in response to altered land area or outgassing rate on WHAK planets. This means smaller changes in precipitation are necessary to alter weathering fluxes to balance land fraction and outgassing changes. Smaller precipitation changes imply smaller temperature changes. 
\subsubsection{Sensitivity to Parameters Controlling Silicate Dissolution Kinetics}
\begin{figure*}[htb!]
    \centering
    \makebox[\textwidth][c]{\includegraphics[width=600pt,keepaspectratio]{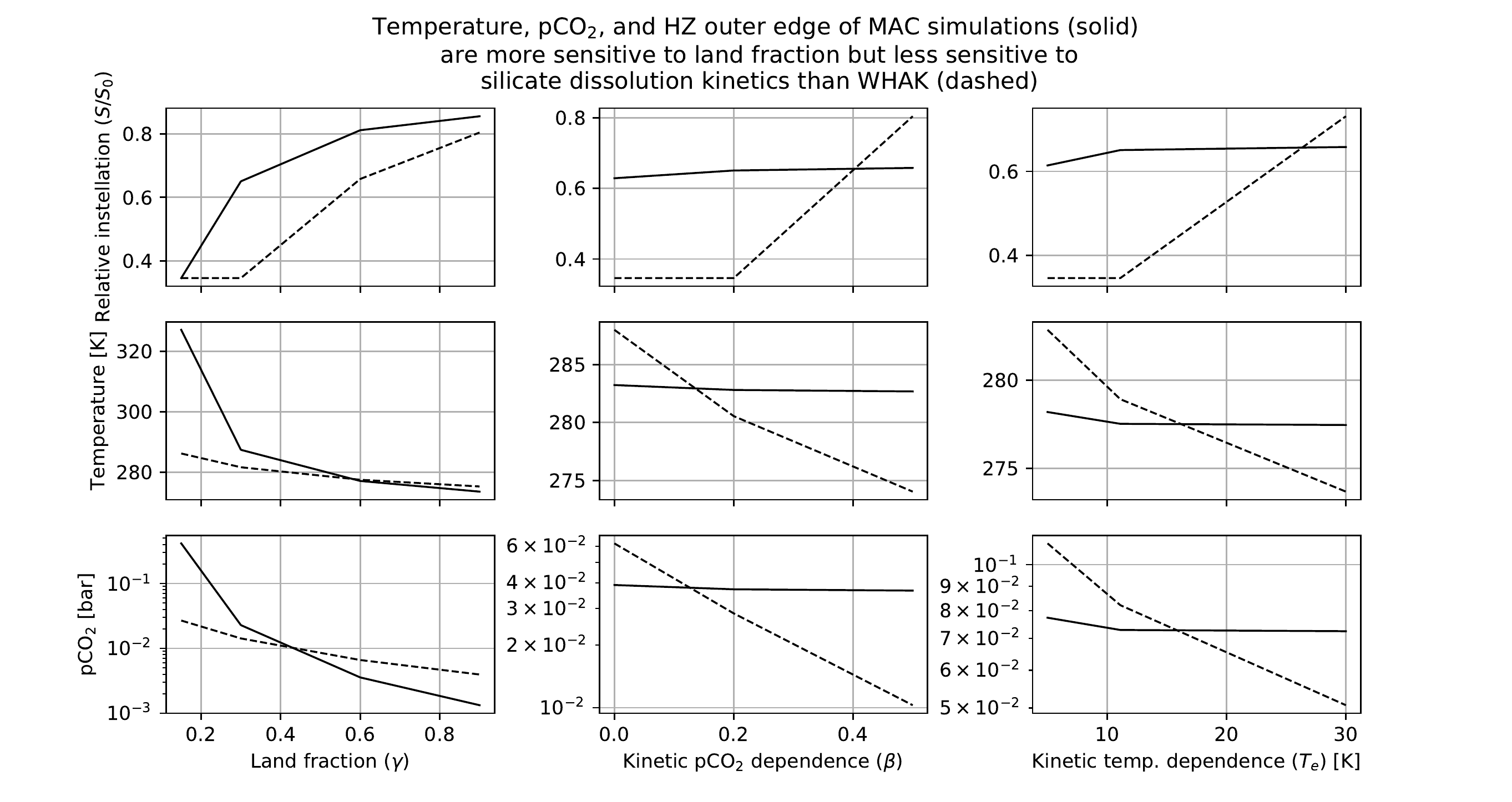}}
    \caption{A comparison of the sensitivity to land fraction and silicate dissolution kinetics of WHAK simulations and MAC simulations. Dashed curves show results for WHAK simulations, solid curve show results for MAC simulations. The top row shows the relative instellation ($S/S_0$, where $S_0=1368$ W m$^{-2}$) of the outer edge of the habitable zone, defined as the instellation where planetary temperature drops below freezing with a given set of weathering parameters. If the calculated outer edge is below the classical outer edge for Earth as defined in \citet{Kopparapu:2013}, the outer edge is set to the classical outer edge ($S/S_0=1/1.7^2=0.346$). The middle row shows the temperature for each simulation at the highest HZ outer edge instellation of the group of simulations in that column. The bottom row shows the pCO$_2$ for each simulation at the highest HZ outer edge instellation of the group of simulations in that column. The leftmost column shows simulations under varying land fraction ($\gamma$), the center column shows simulations under varying kinetic pCO$_2$ dependence ($\beta$), and the right column shows simulations under varying kinetic temperature dependence ($T_e$). Any parameters not being varied take their default value.}
    \label{fig:whak_vs_mac}
\end{figure*}

In agreement with \citet{abbot-2016}, the climates of WHAK planets as a function of instellation in our models are sensitive to the kinetic pCO$_2$-dependence ($\beta$; dashed curves in middle column in Fig. \ref{fig:whak_vs_mac}) and temperature-dependence ($T_e$; dashed curves in rightmost column in Fig. \ref{fig:whak_vs_mac}) in equation \ref{eqn:whak_weather}. For high values of $\beta$, weathering is greatly accelerated by high pCO$_2$, which increases CO$_2$ drawdown and induces cooling at low instellations where high pCO$_2$ is required for habitability. This means that lower $\beta$ values translate to a better climate control as a function of instellation in the WHAK formulation, with the limit of a perfect thermostat at $\beta=0$, other things being equal \citep{Pierrehumbert:2010-book,ramirez2017warmer}. Lower $T_e$ values also translate to more effective climate control, since weathering rates respond more strongly to a given change in temperature.
In contrast to the results for the WHAK formulation, the climates of MAC planets with our default parameter choices show almost no sensitivity to $\beta$ and $T_e$ (solid curves in middle and rightmost columns of Fig. \ref{fig:whak_vs_mac}). This is because increased kinetic dissolution rates (e.g. increased $k_{eff}$) tend to deplete mineral assemblages of reactive components, reducing the impact of kinetics on steady state weathering intensity. Kinetics only influences the steady state weathering flux when the dissolution rate is very slow relative to the rate at which the weathering system is flushed out by runoff. Our MAC simulations are far from the condition where the kinetic dissolution rate determines weathering flux (see final paragraph in Section \ref{subsec:mac}). Instead, weathering flux is determined by the interplay between runoff, cation concentration, and pCO$_2$ governed by the thermodynamics of silicate dissolution and clay precipitation (see Figure \ref{fig:concentration_flux} and Section \ref{subsec:mac}).

\subsection{Sensitivity to Hydrology, Thermodynamic pCO$_2$-dependence, Surface Properties, and Rayleigh Scattering Albedo}

The climates of planets under the MAC formulation are very sensitive to hydrology, the thermodynamics of silicate dissolution and clay precipitation, and surface properties in weathering assemblages because these factors combine to control the concentration and flux of cations to the ocean. Rayleigh scattering albedo can also have a strong effect on the behavior of planets near the outer edges of the habitable zones of G-stars, because of both its direct radiative effects and its effects on weathering arising from energetic limits on precipitation and runoff.

\subsubsection{Hydrology}
The hydrologic cycle is parameterized roughly in our model by making runoff linearly dependent on precipitation and precipitation linearly dependent on temperature. 

\begin{figure*}[htb!]
    \centering
    \makebox[\textwidth][c]{\includegraphics[width=600pt,keepaspectratio]{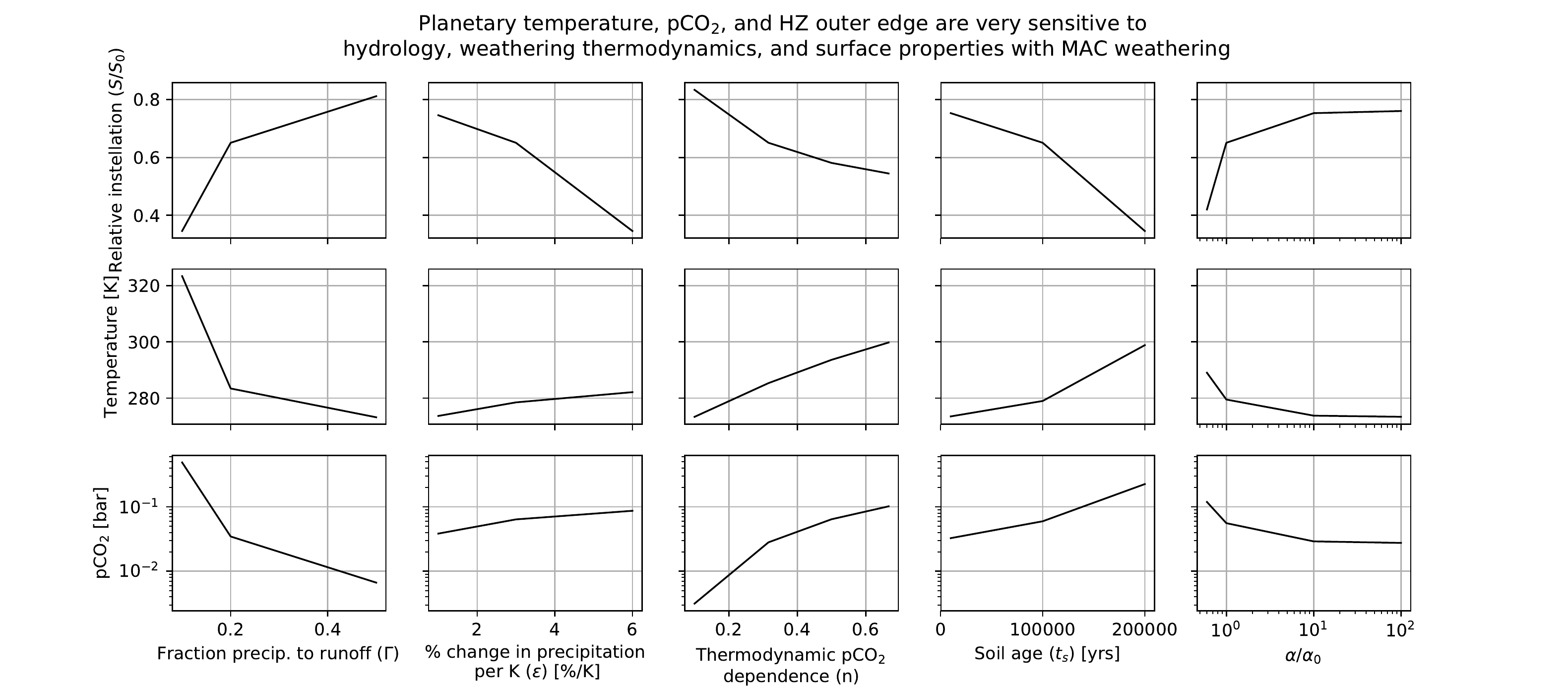}}
    \caption{The sensitivity of MAC weathering simulations to variations in hydrological properties, thermodynamic pCO$_2$ dependence of C$_{eq}$, and surface properties. The top row shows the relative instellation ($S/S_0$, where $S_0=1368$ W m$^{-2}$) of the outer edge of the habitable zone, defined as the instellation where planetary temperature drops below freezing with a given set of weathering parameters. If the calculated outer edge is below the classical outer edge for Earth as defined in \citet{Kopparapu:2013}, the outer edge is set to the classical outer edge ($S/S_0=1/1.7^2=0.346$). The middle row shows the temperature for each simulation at the highest HZ outer edge instellation of the group of simulations in that column. The bottom row shows the pCO$_2$ for each simulation at the highest HZ outer edge instellation of the group of simulations in that column. The leftmost column shows simulations varying the fraction of precipitation that ends up as runoff to the ocean ($\Gamma$), the second column shows simulations varying the $\%$ change in precipitation per Kelvin temperature change ($\epsilon$), the third column shows simulations varying the exponent in the thermodynamic dependence of $C_{eq}$ on pCO$_2$ ($n$), the fourth column shows simulations varying the soil age ($t_s$), and the fifth and final column shows simulations varying the ratio of $L\phi\rho_{sf}AX_r\mu$ to its default value $(\alpha/\alpha_0)$, which can be thought of as the baseline weathering potential for a given mineral assemblage. Any parameters not being varied take their default value.}
    \label{fig:mac_sensitivity}
\end{figure*}

Increasing $\Gamma$, the fraction of precipitation converted to runoff, reduces temperature and pCO$_2$ and moves the effective outer edge of the HZ closer to the star (see the leftmost column of Fig. \ref{fig:mac_sensitivity}). This is because a greater quantity of runoff per unit precipitation leads to a larger weathering flux at a given temperature and pCO$_2$, allowing for a given outgassing rate to be balanced with less rain at colder temperatures. This effect is also illustrated in the bottom panel of Fig. \ref{fig:concentration_flux}: a given weathering flux can be achieved at successively lower pCO$_2$ by increasing runoff (moving rightward in the plot). 
Increasing $\epsilon$, the fractional change in precipitation per unit K, moves the effective outer edge of the HZ away from the star quite strongly, but doesn't have as much impact on temperature and pCO$_2$ (see second column in Fig. \ref{fig:mac_sensitivity}). $\epsilon$ governs the strength of the temperature-dependence of the silicate weathering feedback in the MAC model, since it controls how much the global weathering flux changes in response to temperature changes. Increasing $\epsilon$ makes the weathering flux more sensitive to temperature, which allows a planet with a given set of parameters to maintain balance against CO$_2$ outgassing out to lower instellations before hitting the freezing point. 
\subsubsection{Thermodynamic Control of $C_{eq}$}

Increasing the thermodynamic pCO$_2$ dependence of $C_{eq}$ in isolation ($n$ in equations \ref{eqn:Ceq} and \ref{eqn:mac_weather}) at a given pCO$_2$ below 1 bar tends to reduce $C_{eq}$, which reduces weathering flux, necessitating larger pCO$_2$ to balance outgassing. So, increasing $n$ tends to move the outer edge of the HZ farther from the star and increase temperatures and pCO$_2$ (see middle column of Fig. \ref{fig:mac_sensitivity}).

\begin{figure*}[htb!]
    \centering
    \makebox[\textwidth][c]{\includegraphics[width=500pt,keepaspectratio]{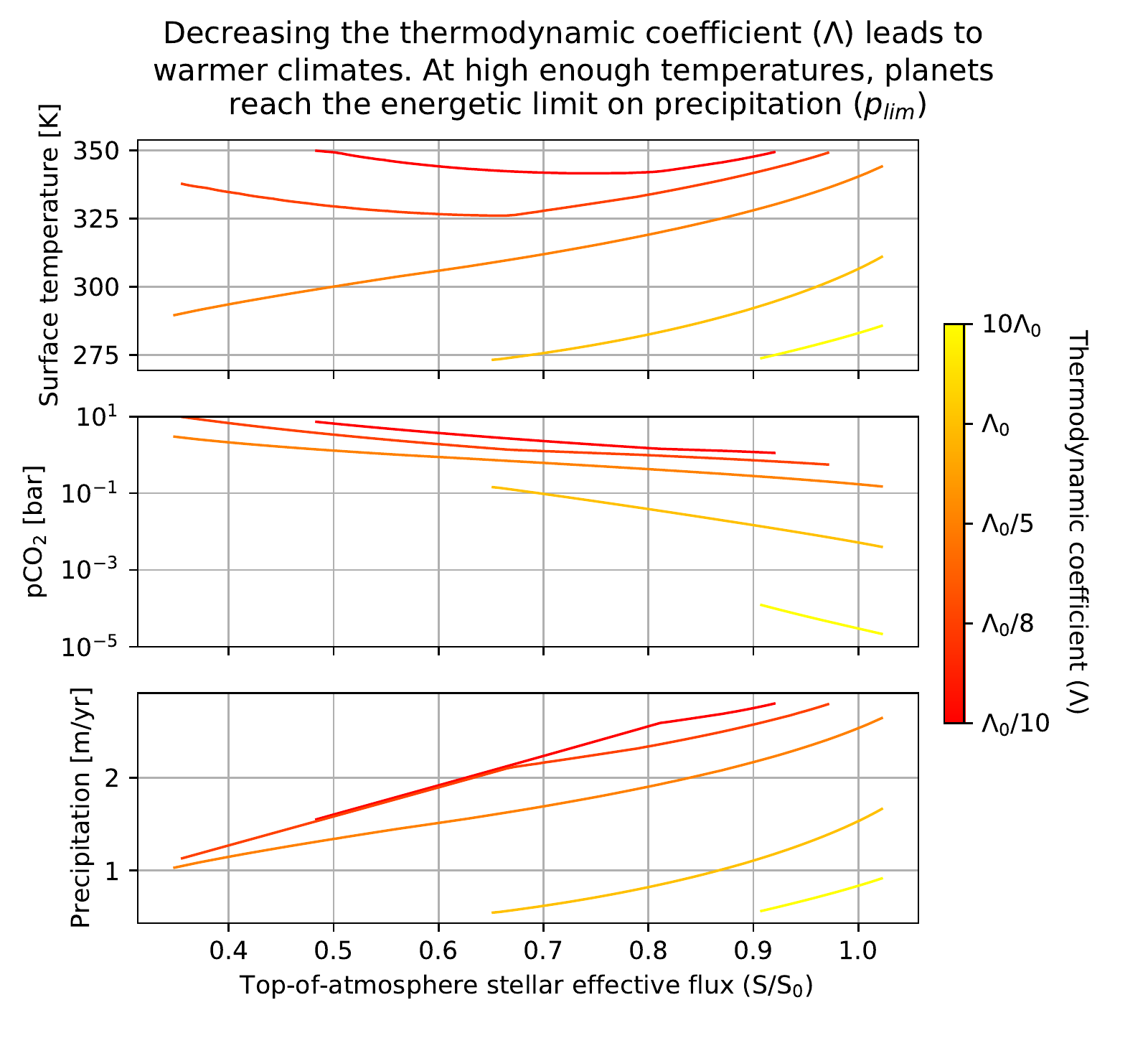}}
    \caption{A comparison of temperature, pCO$_2$, and precipitation vs. instellation for simulations varying thermodynamic coefficient for $C_{eq}$'s dependence on pCO$_2$ ($\Lambda$). The top plot shows instellation vs. temperature ($T$), the center plot shows instellation vs. pCO$_2$, and the bottom plot shows instellation vs. precipitation. The reddest curves use $\Lambda=\Lambda_0/10$ (where $\Lambda_0 = 1.4\times10^{-3}$), and the curves become more yellow as $\Lambda$ is increased through $\Lambda_0/8$,$\Lambda_0/5$, $\Lambda_0$, and $10\Lambda_0$. Parameters not being varied take their default values.}
    \label{fig:LambdaSensitivity}
\end{figure*}

Increasing the thermodynamic coefficient $\Lambda$ tends to increase $C_{eq}$, which allows the concentration of solute in runoff to reach higher levels and therefore allows for a larger maximum weathering flux at a given runoff rate. This means increases to $\Lambda$ cool planets, moving the effective outer edge of the habitable zone closer to the star (Fig. \ref{fig:LambdaSensitivity}).

\subsubsection{Surface Properties}
The physical properties of the surface of a planet have important impacts on its weathering behavior which have not been considered in previous studies of exoplanet weathering. 

Increasing soil age ($t_s$) tends to move the effective outer edge of the HZ farther from the star and increase planetary temperature and pCO$_2$ (see the second-from-right column in Fig. \ref{fig:mac_sensitivity}). This is because older soils are more depleted of weatherable minerals. This increases the time necessary for water moving through a mineral assemblage to reach thermodynamic equilibrium between silicate dissolution and clay precipitation ($t_{eq}$, see Section \ref{subsec:mac}), which decreases solute concentration and increases the quantity of runoff necessary to generate a large enough weathering flux to balance outgassing. At high enough $t_s$, reactive minerals are so depleted that the negative feedback between weathering and climate is lost, as weathering rates are dictated by the supply of reactive minerals to the assemblage (``supply-limited'' weathering \citep{west2005tectonic,foley2015role}--not shown, since no equilibrium climate is possible in this regime with our models). The $t_s$ where supply limitation sets in varies depending on the values taken by other parameters. 

$\alpha$ is the product of several different parameters: reactive flow path length ($L$), porosity of the mineral assemblage being weathered ($\rho$), the ratio of mineral mass to fluid volume ($\rho_{sf}$), the specific surface area of minerals being weathered ($A$), the concentration of reactive minerals in fresh, unweathered bedrock ($X_r$), and a scaling constant that brings the theoretical scaling of solute concentration in line with reactive transport modeling ($\mu$; see the supplement to \citet{maher2014hydrologic}). Increasing any of these parameters brings water flowing through a mineral assemblage closer to thermodynamic equilibrium at a given runoff flux, so $\alpha$ might be thought of as a set of physical properties that determines the baseline weathering potential for a mineral assemblage (an aspect of ``weatherability'', e.g. \citet{kump1997global}). Moving the solute concentration closer to its thermodynamic limit $C_{eq}$ at constant runoff increases the weathering flux (see Fig. \ref{fig:concentration_flux}), drawing down CO$_2$, reducing temperature, and moving the effective outer edge of the habitable zone closer to the star (see rightmost column in Fig. \ref{fig:mac_sensitivity}). 

\subsubsection{Rayleigh Scattering Albedo}
\begin{figure*}[htb!]
    \centering
    \makebox[\textwidth][c]{\includegraphics[width=500pt,keepaspectratio]{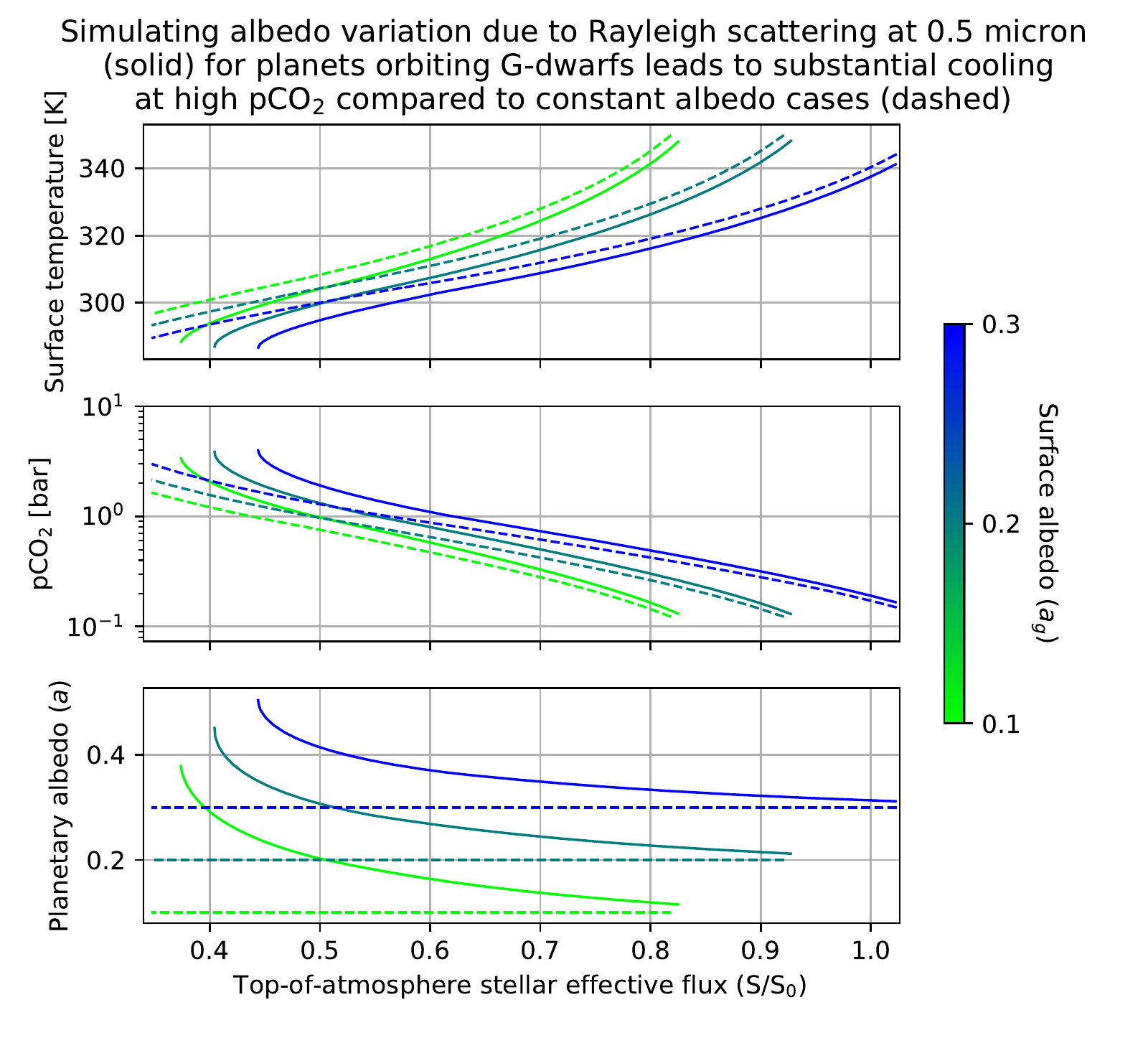}}
    \caption{Comparison of temperature, pCO$_2$, and planetary albedo vs. instellation in MAC simulations with (solid curves) and without (dashed curves) planetary albedo that varies as a function of pCO$_2$ and surface albedo. The top plot shows relative instellation -- the ratio of top-of-atmosphere instellation to modern day Earth instellation ($S/S_0$, where $S_0=1368$ W m$^{-2}$)-- vs. temperature ($T$). The middle plot shows relative instellation vs. pCO$_2$. The bottom plot shows relative instellation vs. planetary albedo. The bluest curves in each plot have a surface albedo $a_g=0.3$, the blue-green curves have $a_g=0.2$, and the green curves have $a_g=0.1$. The thermodynamic coefficient $\Lambda=\Lambda_0/5$ in these simulations so that climates would reach high enough pCO$_2$ for Rayleigh scattering to become important. Other parameters take their default values.}
    \label{fig:alb_sensitivity}
\end{figure*}

We use equation \ref{eqn:planetary_albedo} to estimate the effect on climate of planetary albedo ($a$) variations due to Rayleigh scattering at 0.5 micron by CO$_2$ for Earth-mass planets with a range of surface albedos orbiting Sun-like stars, and we compare the simulations with interactive albedo to simulations with a constant planetary albedo equal to the reflectivity of the surface/clouds (see Fig. \ref{fig:alb_sensitivity}). Our approach to estimating Rayleigh scattering albedo likely overestimates its importance to planetary energy balance, as atmospheric near-infrared absorption should also increase with pCO$_2$, so these simple calculations represent a conservative upper bound on the strength of Rayleigh scattering. We used a reduced thermodynamic coefficient of $\Lambda=\Lambda_0/5$ to decrease $C_{eq}$ and force the simulations to accumulate large pCO$_2$ so that the CO$_2$-induced Rayleigh scattering effect would be clearer. We should also note that any change to a parameter that increases pCO$_2$ (e.g. an increase in soil age) would have the same effect on Rayleigh scattering as the reduction in $\Lambda$ that we use; there is no specific relationship between $\Lambda$ and Rayleigh scattering. 
At low pCO$_2$, the impact of CO$_2$-induced Rayleigh scattering is negligible, leading to changes in albedo of order 1 percent; under these circumstances, planetary albedo is dominated by the prescribed albedo of the surface (which represents the combined surface albedo, cloud albedo, and albedo of the non-CO$_2$ part of the atmosphere). But, as expected, as pCO$_2$ grows with decreasing instellation (middle panel in Fig. \ref{fig:alb_sensitivity}), elevated Rayleigh scattering leads to increased albedo (solid curves in bottom panel of  Fig. \ref{fig:alb_sensitivity}), which in turn cools a simulation relative to the corresponding simulation without Rayleigh scattering (compare solid and dashed curves in top panel of Fig. \ref{fig:alb_sensitivity}). This cooling effect moves the effective outer edge of the habitable zone closer to the star. 

\subsection{Impact of an Energetic Limit on Precipitation} \label{subsec:energetic_limit}
In steady-state, precipitation is in equilibrium with evaporation, which is ultimately driven by instellation. This implies a maximum planetary precipitation rate constrained by planetary instellation (equation \ref{eqn:plim}), a prediction which is borne out by both 1-dimensional \citep{Pierrehumbert-2002:hydrologic} and 3-dimensional \citep{o2008hydrological,LeHir:2009p898} climate models. 
\subsubsection{Energetic Limit Without Land Fraction Dependence}
The impact of the energetic limit on precipitation is illustrated in Figure \ref{fig:LambdaSensitivity}. Reducing $\Lambda$ (the thermodynamic coefficient in equation \ref{eqn:Ceq}) reduces $C_{eq}$ for a given pCO$_2$, which increases the temperature, pCO$_2$, and precipitation necessary to achieve a runoff high enough to balance outgassing. Simulations with $\Lambda=\Lambda_0/5$ and above (where $\Lambda_0$ is the default value for $\Lambda$ listed in Table \ref{tab:values}) have temperature slopes identical to those of their precipitation curves because precipitation is linear with respect to temperature. However, for the simulations with $\Lambda_0/8$ and $\Lambda_0/10$, temperature eventually becomes decoupled from precipitation and begins to increase instead of decrease as instellation falls. This is because the temperatures in these simulations are forced so high by the restriction of $C_{eq}$ that they each achieve the maximum precipitation at a given instellation somewhere in the HZ. For simulations that reach their maximum precipitation rate at some instellation, precipitation then follows $p_{lim}$ and decreases linearly with instellation beyond that (note that the red and red-orange precipitation curves converge onto the same line in the bottom panel of Fig. \ref{fig:LambdaSensitivity}, since they both follow equation \ref{eqn:plim}). Since precipitation (and therefore runoff) decreases faster with instellation when governed by equation \ref{eqn:plim}, this forces pCO$_2$ to grow at a higher rate with decreasing instellation in order to increase $C_{eq}$ and allow solute to accumulate to high enough concentrations to maintain a weathering flux balanced with outgassing despite reduced runoff (see the bottom panel of Fig. \ref{fig:concentration_flux} to see the relationship between pCO$_2$, runoff, and weathering flux for a given set of parameters). This increased CO$_2$ accumulation per unit instellation reduction is what leads to the counter-intuitive temperature increase at low instellations in the red and red-orange curves in the top panel of Fig. \ref{fig:LambdaSensitivity}.  

\subsubsection{Energetic Limit With Land Fraction Dependence}
\begin{figure*}[htb!]
    \centering
    \makebox[\textwidth][c]{\includegraphics[width=400pt,keepaspectratio]{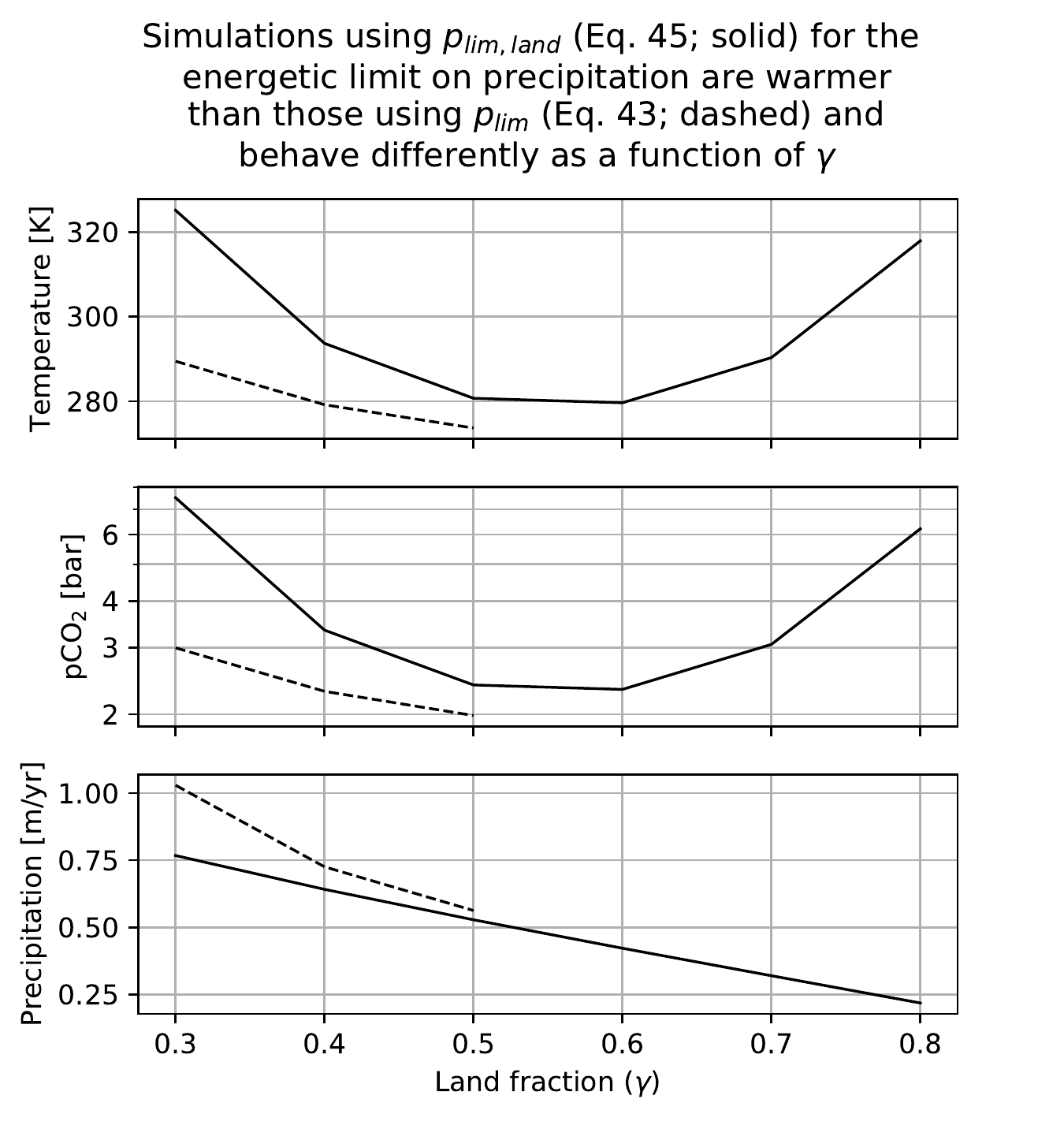}}
    \caption{A comparison of temperature, pCO$_2$, and precipitation at the classical outer edge of Earth's habitable zone ($S/S_0=0.346$ \citep{Kopparapu:2013}) as a function of land fraction ($\gamma$) for planets with an energetic limit on precipitation that either scales with ocean surface fraction (solid curves; equation \ref{eqn:plimland}) or is independent of ocean surface fraction (dashed curves; equation \ref{eqn:plim}). The top panel shows $\gamma$ vs. temperature ($T$), the middle panel shows $\gamma$ vs. pCO$_2$, and the bottom panel shows $\gamma$ vs. precipitation ($p$). The simulations have thermodynamic coefficient $\Lambda=\Lambda_0/5$, but other parameters take their default values.}
    \label{fig:plimland}
\end{figure*}

We also ran a set of simulations with $p_{lim,land}$ (equation \ref{eqn:plimland}) instead of $p_{lim}$ (equation \ref{eqn:plim}) to examine the impact of scaling the energetic limit on global precipitation by surface ocean fraction (1-$\gamma$). This leads to several changes in the weathering behavior of planets in the energetically limited regime as a function of land fraction $\gamma$. This is demonstrated in Figure \ref{fig:plimland}, which compares the temperature, pCO$_2$, and precipitation for simulations with thermodynamic coefficient $\Lambda=\Lambda_0/5$ at $S/S_0=0.346$ (the classical outer edge of the habitable zone for an Earth-mass planet around a G2-star according to \citet{Kopparapu:2013}) with and without the new scaling law for the energetic limit. 

The first point to note is that scaling the energetic limit with ocean fraction lowers the temperature at which a planet's precipitation becomes energetically limited for a given instellation, since a reduced fraction of the instellation is available to drive precipitation. For the simulations that use $p_{lim}$, a thermodynamic coefficient of $\Lambda_0/5$ (with other parameters set to default values) leads to a surface temperature that is not high enough to enter the energetically-limited regime at any instellation (see Fig. \ref{fig:LambdaSensitivity} and the dashed curves in Fig. \ref{fig:plimland}) regardless of land fraction; in contrast, simulations that use $p_{lim,land}$ instead of $p_{lim}$ enter the energetically limited regime for all $\gamma$ when $\Lambda=\Lambda_0/5$ (see solid curves in Fig. \ref{fig:plimland}).
Another difference between these sets of simulations is that energetically-limited simulations that use $p_{lim,land}$ tend to display warmer temperatures and higher pCO$_2$ than those using $p_{lim}$ (see top and middle panels of Fig. \ref{fig:plimland}). This is because simulations with $p_{lim,land}$ are restricted to a lower maximum precipitation rate (see bottom panel in Fig. \ref{fig:plimland}). A lower precipitation rate implies a lower rate of runoff in our models, which forces pCO$_2$ to build to higher levels in order to increase the thermodynamic limit $C_{eq}$ enough for the low runoff to deliver sufficient cations to the ocean to balance outgassing of CO$_2$.

Finally, there is a qualitative difference in the dependence of planetary climate and weathering behavior on land fraction ($\gamma$) for energetically-limited planets with $p_{lim,land}$. In our simulations, under non-energetically-limited conditions (and under energetically-limited conditions governed by $p_{lim}$ instead of $p_{lim,land}$), increased land fraction leads to cooling  (e.g. Fig. \ref{fig:land_frac_whak_mac} or the leftmost column of Fig. \ref{fig:whak_vs_mac}). This is because a larger land area allows a given outgassing rate to be balanced with a smaller weathering flux per unit land surface, necessitating lower precipitation rates and temperatures. In contrast, energetically-limited simulations with $p_{lim,land}$ may warm or cool with increased $\gamma$ (see top panel of Fig. \ref{fig:plimland}). When cooling dominates, it is because of the effect just discussed. When warming occurs with increased land fraction, it is because $p_{lim,land}$ (and therefore runoff) becomes very small at high $\gamma$ since the ocean surface fraction becomes small, which forces pCO$_2$ to increase greatly so that $C_{eq}$ becomes large enough that the tiny rates of runoff can still deliver enough cations to the ocean to balance outgassing.

\section{Discussion $\&$ Conclusions}\label{sec:discussion}

\subsection{Discussion}

\subsubsection{The Influence of Tectonic Mode on Silicate Weathering}
In this study, we did not address the tectonic regime of the planets we modeled, but weathering is probably strongly impacted by tectonic mode. Planets with plate tectonics have a robust mechanism for removing weathered material and delivering fresh weatherable material to the planetary surface, potentially improving their ability to avoid a ``supply-limited'' weathering regime where the negative feedback between climate and weathering is lost due to depletion of weatherable minerals \citep{west2005tectonic,foley2015role}. Simulations of plate tectonic planets display efficient CO$_2$ degassing under most conditions \citep{noack2014can}. Simulations of stagnant lid planets suggest that they may also be able to avoid the supply limit over geologic timescales \citep{foley2018carbon,foley2019habitability} and display efficient degassing and volcanism under a range of conditions, though their volcanic activity may become limited at high core-mass fraction or high planetary mass \citep{noack2014can,noack2017volcanism,dorn2018outgassing}. Between these endmembers, there are intermediate states like the ``episodic'' mode, with long periods of stagnant lid behavior punctuated by periods of relatively rapid resurfacing and outgassing \citep{lenardic2016climate}.

The MAC formulation suggests some natural ways to explicitly incorporate tectonic mode into simulations of planetary weathering behavior. Through its impact on the Damk\"{o}hler coefficient ($D_w$), the average soil age of material being weathered in our simulations ($t_s$) has a powerful impact on the weathering behavior of rocky planets. Young soils lead to large values of $D_w$, allowing for a robust weathering feedback. Ancient soils that are depleted in reactive minerals can enter a supply-limited weathering regime. We would expect that planets with more rapid resurfacing will, on average, have younger soils and therefore greater values of $D_w$. They should also have higher volcanic outgassing rates. Younger soils lead to more efficient weathering and therefore lower equilibrium pCO$_2$ and temperatures. Higher outgassing rates lead to higher equilibrium pCO$_2$ and temperature. Since greater volcanic activity may lead to both faster resurfacing and higher rates of outgassing, the effects tend to offset one another and it is difficult to predict what the net impact on planetary weathering behavior would be.

Tectonic mode should also influence the lithology of the material being weathered on a given planet. For example, the relative abundances of granitic and basaltic weatherable material may be determined by the relative rates of continental generation vs. production of flood basalts, which will in turn be controlled by the primary mode of resurfacing on a planet. \citet{ibarra2016differential} showed that granitic and basaltic river catchments on Earth display divergent weathering behavior: rivers draining basaltic catchments tend to have higher values of $C_{eq}$ and cation concentration. \citet{winnick2018relationships} argued that the weathering of basaltic minerals should display greater sensitivity to pCO$_2$ because basalts have a higher proportion of divalent cations, which leads to a larger value for the thermodynamic pCO$_2$ sensitivity ($n$; see equations \ref{eqn:n_calc} and \ref{eqn:Ceq}). Those results imply that planetary weathering dominated by granites vs. basalts could lead to quite different equilibrium climate states for planets with otherwise similar properties.

We also note the importance of mountain building processes for the silicate weathering feedback. Much of the chemical weathering on Earth takes place in mountainous regions where fresh minerals are exposed to the Earth's surface and soil is produced and eroded at high rates \citep{larsen2014rapid,larsen2014contribution}. Since orogeny is heterogeneous in space and time on Earth, the build-up and grind-down of mountains, as well as the climatic conditions at the locations on Earth where these processes take place, may have a profound impact on the carbon cycle and global climate of a planet through time. Periods of active mountain-building (particularly in areas with appreciable runoff) may lead to global cooling by increasing the fraction of fresh minerals ($f_w$; equivalent to reducing soil age) in the region being weathered, whereas periods of slow uplift may allow CO$_2$ to accumulate to high levels in the atmosphere \citep{raymo1992tectonic,jagoutz2016low,kump2018prolonged,macdonald2019arc}

\subsubsection{Relation to Previous Rocky Exoplanet Weathering Results}
To our knowledge, all previous rocky exoplanet weathering studies have used some variant of the WHAK formulation to represent weathering. Simulations with the MAC formulation display some important differences in behavior compared to those with the WHAK formulation. So, if we assume that the MAC formulation is an accurate representation of planetary silicate weathering, some conclusions of these previous studies should be revisited. We survey a few points of interest below.

As we noted in Section \ref{sec:results}, \citet{abbot12-weathering} and \citet{foley2015role} found that planetary weathering behavior is relatively insensitive to land fraction with the WHAK model. We found the opposite with the MAC model: land fraction strongly impacts the planetary temperature, particularly at low land fraction. This suggests that understanding the processes that control oceanic volume \citep[e.g.][]{Kasting1992,Cowan-2014,schaefer2015persistence,komacek2016effect} and the development of continents \citep[e.g.][]{rosing2006rise,honing2014biotic,honing2016continental,honing2019bifurcation} on exoplanets is extremely important for predicting the frequency of occurrence of truly habitable planets.

Snowball limit cycling is a hypothetical phenomenon on terrestrial planets where silicate weathering draws down CO$_2$ faster than outgassing supplies it, even at temperatures cold enough to freeze a planet, plunging a planet into the snowball state. In the snowball state, silicate weathering is expected to slow down or cease, allowing CO$_2$ to outgas and build up until a planet has warmed enough to deglaciate, at which point rapid silicate weathering freezes it again \citep{Tajika2007,Menou2015,batalha2016climate,haqq2016limit,abbot-2016,paradise2017,ramirez2017warmer} (although CO$_2$ condensation in the extreme cold of the snowballs or continued weathering at the seafloor during the snowball may make deglaciation difficult under some conditions \citep{turbet2017co,kadoya2019outer}). In simulations with the WHAK formulation, limit cycling is driven by the kinetic dependence of weathering on pCO$_2$ ($\beta$) increasing weathering rates at low instellations where pCO$_2$ is very high. In the MAC formulation, kinetic pCO$_2$-dependence has essentially no impact on planetary weathering rates (see the middle column of Fig. \ref{fig:whak_vs_mac}), but the thermodynamic dependence of $C_{eq}$ on pCO$_2$ leads to a similar result. Many parameter configurations lead to planets with equilibrium temperatures below freezing at instellations within the habitable zone, so planets with these parameter sets at low enough instellations would either be locked into a permanent snowball state or go through limit cycling. The fact that factors like hydrology and soil age and lithology impact the effective outer edge of the habitable zone so strongly (see top row in Fig. \ref{fig:mac_sensitivity}) implies that a planet's susceptibility to limit cycling (and therefore its habitability) is determined by a complex interplay of factors that are not easily constrained a priori. Arguing that a given planet (e.g. Mars \citep{batalha2016climate}) experienced limit cycling thus requires making many implicit and potentially unfounded assumptions about the properties that controlled its weathering (see \citet{ramirez2017warmer,hayworth2020warming} for further discussion of the early Martian limit cycling hypothesis).


\citet{Kite:2011} use the WHAK formulation of weathering to argue that there may be conditions where silicate weathering acts as a destabilizing feedback on the climates of rocky tidally-locked planets. The ``enhanced substellar weathering instability'' (ESWI) is proposed to take place on tidally locked planets with relatively thin atmospheres composed largely of CO$_2$. Because heat transport in thin atmospheres decreases with reductions in density, a decrease in atmospheric mass can lead to net warming at the substellar point on a tidally locked planet despite the reduction in greenhouse effect, due to decreased ability of the atmosphere to export solar energy away from the substellar point. Because of the exponential dependence of weathering rates on temperature in the WHAK formulation, the vast majority of weathering on a tidally locked planet occurs near the substellar point, so an increase in temperature at the substellar point through a reduction in atmospheric pressure can enhance global weathering significantly. The enhanced weathering would in turn draw down more CO$_2$, again reducing atmospheric pressure, warming the substellar point, and enhancing weathering. This is a positive feedback leading to atmospheric collapse, since weathering begets more weathering, and the reverse scenario of runaway CO$_2$ accumulation also occurs in the case of an initial increase in atmospheric mass (or an initial cooling). However, the ESWI depends intimately on the strength of the temperature-dependence of the silicate weathering feedback. The non-exponential temperature-dependence in the MAC formulation might lead to global weathering that is less concentrated at the substellar point, implying that tidally locked planets with thin CO$_2$-dominated atmospheres may be less vulnerable to the ESWI than suggested by \citet{Kite:2011}. 

\textcolor{black}{\subsubsection{Implications for Early Earth Climate Evolution}
During Earth's Archean eon (4 to 2.5 Ga), the Sun was likely 20-30$\%$ fainter than today \citep[e.g.][]{SAGAN:1972p1233,bahcall2001solar}. Sparse proxy estimates from this period seem to imply a temperate world with ocean temperatures $<40$\degree C and occasional partial glaciations \citep[e.g.][]{hren2009oxygen,blake2010phosphate,ojakangas2014talya,de20163,catling2020archean}, suggesting the climate was effectively buffered against reduced luminosity. The question of how the Earth maintained habitability despite this large change in luminosity is sometimes referred to as the ``Faint Young Sun problem,'' a particularly famous component of the more general problem of constraining the evolution of the climate during Earth's deep prehistory. As we noted in the introduction, the silicate weathering feedback is often invoked as a source of the climatic buffering in Earth's past, and this interpretation is consistent with carbon cycle models of the Archean period that use WHAK-style kinetic representations of continental and seafloor weathering to estimate Archean Earth's equilibrium climate state \citep[e.g.][]{charnay2017warm}. In fact, the proxy record of ancient climate has been combined with WHAK-based inverse geologic carbon cycle modeling in attempts to simultaneously provide tighter constraints on Earth's temperature-CO$_2$ history, key globally-averaged parameters like $T_e$ and $\beta$ in the WHAK formulation, and the relative importance of processes like seafloor weathering and reverse weathering compared to continental weathering in the past  \citep{krissansen2017constraining,krissansen2018constraining,krissansen2020coupled}. These examples illustrate how the WHAK framework, as a central component of most models of the carbon cycle of the ancient Earth, contributes to current understanding of the Archean Earth system and the Faint Young Sun problem.}

\textcolor{black}{As shown in Section \ref{subsec:whakvmac}, the MAC and WHAK formulations of weathering lead to different atmospheric responses to changes in boundary conditions. MAC is much more sensitive to changes in land fraction ($\gamma$) and volcanic outgassing ($v$), displaying a 30-40 K temperature increase in response to a factor-of-2 reduction in $\gamma$/$v$, compared to a $\sim 5$ K increase for WHAK  (see Fig. \ref{fig:land_frac_whak_mac} and the leftmost column of Fig. \ref{fig:whak_vs_mac}, noting that reductions in $v$ are equivalent to proportional increases in $\gamma$). On the other hand, depending on parameter choices like $n$ and $\beta$, either formulation of weathering can be more sensitive to changes in luminosity. On Earth, all of these boundary conditions have changed substantially since the Archean, with a general secular decrease in outgassing by as much as a factor of 10 \citep{avice2017origin} and an increase in dry land by a factor of 10 or more \citep{flament2008case,johnson2020limited} accompanying the increase in solar luminosity already mentioned. This suggests an increase in $\gamma / v$  by perhaps a factor of $\sim$ 100 (!) since the Archean, which is enormous compared to the factor of 2 reduction that causes 30-40 K of warming in our default MAC model. Of course, other parameters in the MAC model have certainly changed since the Archean\textemdash we specifically discuss these boundary conditions because the magnitude of change over time and the divergence in response by WHAK and MAC help illustrate the point that replacing the WHAK formulation in carbon cycle models with a MAC-style parameterization could lead to substantially different conclusions about the evolution of the Earth system since the Archean. }

\textcolor{black}{Since the change in luminosity sensitivity between the two formulations can go in either direction, we will also ignore that potential difference between WHAK and MAC and focus on sensitivity to land fraction and outgassing. Because MAC weathering requires larger increases in temperature than WHAK to maintain equilibrium in response to reduced $\gamma/v$, and because $\gamma/v$ may have been a factor of 100 smaller during the Archean, replacing the WHAK model with the MAC formulation would lead to a much hotter inferred Archean climate, other things being equal. However, as noted in the beginning of this section, the (very limited) climate proxy estimates from the Archean suggest a world with a temperate climate not much warmer than our own. Assuming those estimates are trustworthy, the behavior of the MAC model may make it necessary to invoke other changes to the carbon cycle, e.g. a greater seafloor weathering flux, to generate enough weathering to balance outgassing at temperatures consistent with the mild climate implied by proxies. Other differences, particularly the lack of vascular land plants, may imply further weakening of Archean continental weathering \citep[e.g.][]{rafiei2019weathering}, though the influence of plants is still poorly understood. These suggestions can be tested quantitatively. The implications of the MAC formulation for our interpretation of the proxy record of the evolution of the Earth system should be addressed rigorously with a statistical methodology that accounts for the huge uncertainties in most parameters when modeling the ancient carbon cycle (see the approach taken in \citet{krissansen2017constraining,krissansen2018constraining,krissansen2020coupled}).} 

\subsubsection{Observability of the Silicate Weathering Feedback}\label{subsubsec:observations}
\citet{bean2017,checlair2019statistical} suggest the use of ``statistical comparative planetology'' to detect the operation of the silicate weathering feedback on rocky planets in the HZ. Specifically, \citet{bean2017,checlair2019statistical} propose using a relatively large number of low cost, low precision CO$_2$-abundance measurements of rocky planets throughout the habitable zone with a future telescope to show that pCO$_2$ tends to decrease with increases in instellation, which would provide evidence for a stabilizing feedback on planetary climate. With the assumption that the silicate weathering feedback would adjust pCO$_2$ to hold planetary temperatures at 280 K, and with very optimistic assumptions for observational noise, \citet{checlair2019statistical} conclude that $\sim10$ planets with functioning silicate weathering feedbacks in the habitable zone would need to have their CO$_2$ partial pressures characterized to give an 80$\%$ chance of detecting a trend in pCO$_2$ vs. instellation. Under pessimistic instrumental noise assumptions, \citet{checlair2019statistical} calculate that the number of necessary characterizations increases to $\sim50$. These numbers bracket the range of estimated yields for ``exo-Earth'' planets from proposed next-generation telescopes HabEx \citep{gaudi2020habitable} and LUVOIR \citep{luvoir2019luvoir}. However, it is worth noting that recent studies suggest that the occurrence rate of Earth-like planets may be up to an order of magnitude lower than the numbers used to calculate those estimates \citep{pascucci2019impact,neil2019joint}, which would lead to a proportional reduction in yield.

Our study suggests a larger number of observations may be required to establish a trend of pCO$_2$ vs. instellation and confirm the operation of the silicate weathering feedback, due to the huge variations in pCO$_2$ and temperature at a given instellation for planets with different different outgassing rates or combinations of hydrological, thermodynamic, and surface mineralogical parameters. As an example, see the middle panel in Fig. \ref{fig:LambdaSensitivity}: at $S/S_0\sim0.9$, equilibrium pCO$_2$ varies by 4 orders of magnitude between the simulations with $\Lambda=\Lambda_0/10$ and $10\Lambda_0$. Both of those values of $\Lambda$ are within the range expected from diversity of silicate mineral assemblage lithologies \citep{winnick2018relationships}. Varying other parameters produces a similar spread in pCO$_2$. Effectively, the potential diversity in weathering behavior among terrestrial planets adds a form of intrinsic ``noise'' to the statistical estimates in \citet{checlair2019statistical}. To add to the complication, variations in seafloor weathering or reverse weathering (discussed below in section \ref{subsubsec:limitations}) could also add to the variety of climates encountered in the HZ. 

The diversity in equilibrium climates permitted by variations in hydrological, thermodynamic, and surface parameters, coupled with the potentially low occurrence rate of Earth-like planets noted above, suggests that the method outlined in \citet{bean2017,checlair2019statistical} may not be adequate to detect the silicate weathering feedback with upcoming telescope missions. Because of this, we suggest that a complementary method first suggested in \citet{turbet2019two} to identify the breakdown of the silicate weathering feedback at the inner edge of the habitable zone may be able to successfully demonstrate the existence of the feedback with a smaller number of observations. We do not statistically quantify the number of characterizations necessary to identify the operation of the silicate weathering feedback with this method, as that is beyond the scope of the current study.
\begin{figure*}[htb!]
    \centering
    \makebox[\textwidth][c]{\includegraphics[width=500pt,keepaspectratio]{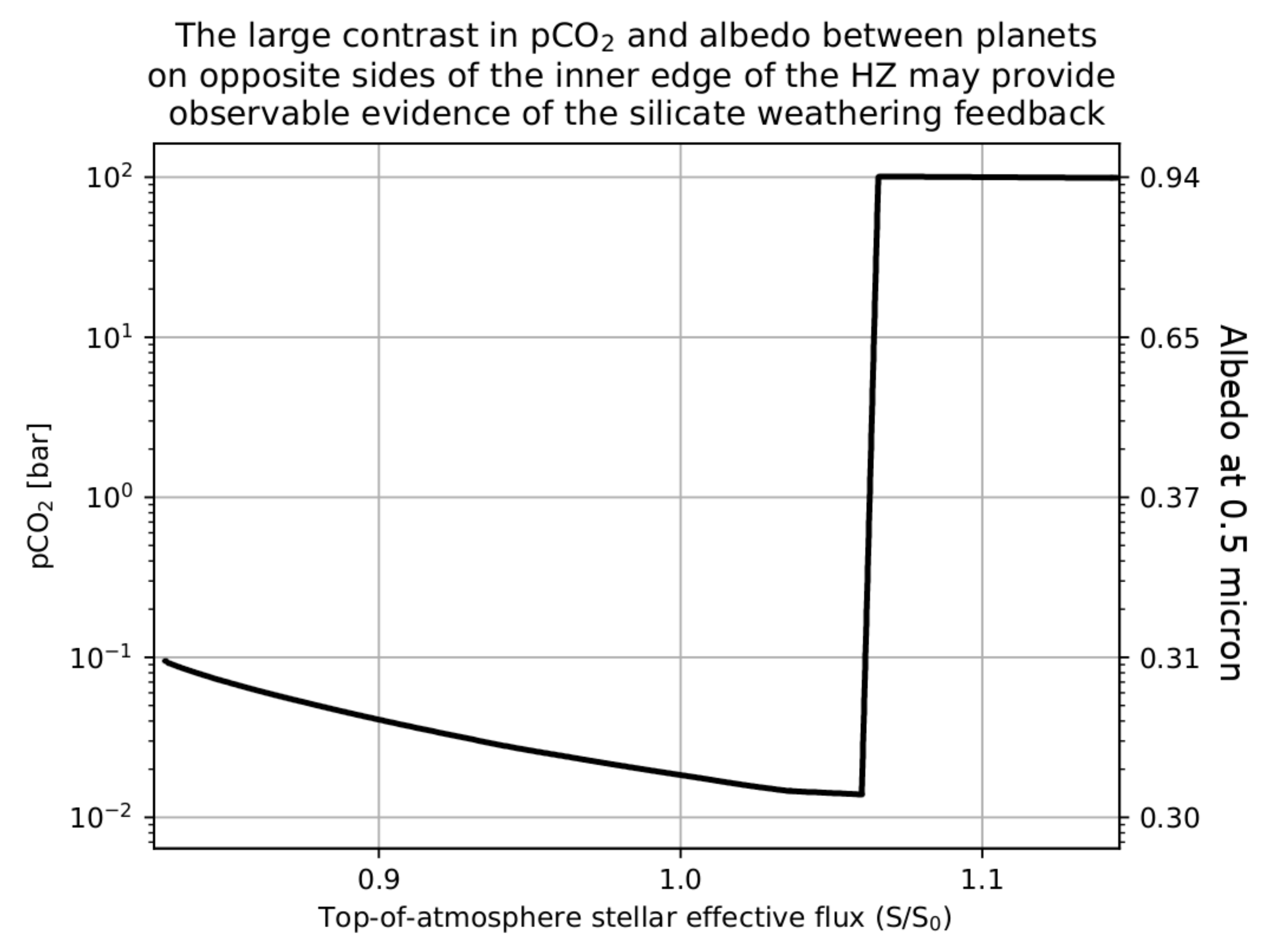}}
    \caption{A schematic plot showing variation of pCO$_2$ (left y-axis) and albedo (right y-axis) with instellation across the inner edge of the habitable zone (defined as the instellation at which a planet enters a runaway greenhouse and loses its oceans to space). This curve is an illustration of what we expect to see based on the contrast between Earth and Venus in our solar system, and does not represent calculations we carried out. Planets with functional silicate weathering feedbacks within the habitable zone but near its inner edge will have low pCO$_2$ due to high instellation, while planets closer to the star than the inner edge of the habitable zone may experience catastrophic decarbonation of their carbonate minerals due to extreme heating by a post-runaway greenhouse steam atmosphere. The orders-of-magnitude difference in the pCO$_2$ and large contrast in albedo on opposite sides of the inner edge of the HZ may allow a statistically robust observation of the ``pCO$_2$ cliff'' with a small number of characterizations of planets' CO$_2$ abundances and/or albedos, confirming the operation of the silicate weathering feedback and helping to better constrain the location of the inner edge. The dependence between pCO$_2$ and albedo we used to generate the albedo axis in this plot assumes a surface albedo of 0.3; with a lower surface albedo, the contrast in albedos across the inner edge of the HZ would be even larger.}
    \label{fig:inner_edge}
\end{figure*}
The method from \citet{turbet2019two} that we will discuss makes use of the potential for a large discontinuity in CO$_2$ concentration between planets on opposite sides of the inner edge of the habitable zone, defined here as the instellation above which a planet enters the runaway greenhouse state and loses its water to space \citep{Kasting93}. There is not a unique instellation where the runaway greenhouse occurs, as the critical irradiation level depends on planetary rotation rate \citep{yang2014}, planetary radius \citep{yang2019effects}, surface gravity \citep{yang2019effects}, stellar temperature \citep{Kopparapu:2013}, background gas partial pressure \citep{ramirez2020effect}, surface water distribution \citep{kodama2018dependence,kodama2019inner}, and cloud properties \citep{leconte2013increased}, but for the purposes of this discussion, we will use a conservative estimate for an Earth-mass, rapidly-rotating planet around a Sun-like star given in \citet{Kopparapu:2013}. In that study, the critical instellation for a planet with those properties is found to be $S_{eff}=1.05$ (where $S_{eff}$=$S/S_{earth}$ is the ratio of flux received by the planet to the flux received by modern Earth). 

For planets just beyond the inner edge (e.g. planets just within the HZ) with a functional silicate weathering feedback, CO$_2$ will be drawn down to low levels due to the high instellation. To give a conservative maximum estimate of pCO$_2$ at the inner edge of the habitable zone, we use eqn. \ref{eqn:olr} to find the pCO$_2$ necessary to equilibrate a planet with an albedo of 0.3 and $S_{eff}=1.05$ at a surface temperature of 340 K. This yields a pCO$_2$ of 0.05 bar. In actuality, the pCO$_2$ could be considerably lower, depending on the various parameters that control the weathering rate, but with a higher pCO$_2$, a planet would be hot enough to enter a moist greenhouse regime and lose its water to space efficiently \citep{kasting1988climate}.

In contrast, Earth-like planets within the inner edge of the HZ that undergo a runaway greenhouse and lose their oceans to space may end up in a ``Venus-like'' state with $\sim100$ bars of CO$_2$ in their atmosphere due to the catastrophic decarbonation of their mineral inventories under extreme temperatures (though this prediction depends on the magnitude of carbon delivery to a planet during its accretion phase). Even in the absence of catastrophic decarbonation, if a planet loses its water due to the runaway but still has even a little bit of CO$_2$ outgassing (and an Earth-like inventory of carbon), a Venus-like CO$_2$-rich atmosphere will accumulate, since a fully dry planet cannot maintain a silicate weathering feedback. This suggests that there could be a 2-3 order of magnitude discontinuity in pCO$_2$ for Earth-like planets straddling opposite sides of the inner edge of the habitable zone, assuming an operative silicate weathering feedback within the HZ.

If present, the multiple-orders-of-magnitude change in pCO$_2$ across the inner edge of the HZ (see Fig. \ref{fig:inner_edge} for a schematic illustration of this concept) may be observable with next-gen telescopes like OST, HabEx, or LUVOIR without the need for a large number of atmospheric characterizations. However, it may be difficult to demonstrate that a planet has CO$_2$ partial pressures greater than a bar or two via infrared emission, since CO$_2$ becomes optically thick throughout the IR region at high column abundances; for example, remote observations of the night-side of Venus look similar to the day-side of Mars in the IR despite a factor of $\sim16000$ difference in CO$_2$ partial pressure, since IR only escapes from the top $\sim$bar of Venus's atmosphere (see Fig. 3 in \citet{Pierrehumbert2011-infrared}). 

In the absence of direct measurements of atmospheric pCO$_2$ at the surface of Venus-like planets, we propose that the contrast in albedo across the inner edge of the HZ at visible wavelengths due to Rayleigh scattering by CO$_2$ should be substantial and likely observable: low pCO$_2$ planets near the inner edge of the habitable zone would have albedos approximately equal to that of their surface + clouds, suggesting a maximum of around $0.3$ (based on Earth's albedo in the visible \citep{tinetti2006detectability}), whereas Venus-like planets would have visible-wavelength albedos $>0.9$ due to efficient Rayleigh scattering by $\sim$100 bar of CO$_2$, even in the absence of shiny sulfuric acid clouds like those hosted by Venus (calculated at 0.5 micron with equation \ref{eqn:planetary_albedo}; see also Table 5.4 in \citet{Pierrehumbert:2010-book}). Note that we are specifically discussing albedo in the visible region of the spectrum: albedo integrated across the spectrum would include near-IR absorption effects, which would reduce the albedo contrast across the inner edge of the HZ and introduce a dependence on stellar type since lower-temperature stars emit preferentially at longer wavelengths. By restricting the discussion to visible-wavelength albedo, we avoid those complications. 

In summary, observations of large differences in pCO$_2$ and/or visible-wavelength albedo between planets on opposite sides of the inner edge of the habitable zone may allow for a demonstration that the silicate weathering feedback stabilizes planetary climate within the HZ and fails to stabilize climate outside of the HZ. In particular, because of the abrupt and large change in pCO$_2$ and albedo across the inner edge of the HZ, the existence of the silicate weathering feedback may be demonstrable with a smaller number of observations than a method that depends on the potentially noisy trend of pCO$_2$ vs. irradiation due to silicate weathering within the habitable zone. 

\textcolor{black}{\subsubsection{Relating Planetary Characteristics to Weathering Controls}
The potential diversity in climate outcomes we discuss above partially might partially from our ignorance of the constraints on parameters controlling weathering in the MAC formulation. A better understanding of the likely distributions of important variables like soil thickness and age, particularly as functions of bulk parameters like planetary mass, might serve to narrow the range of climate states we expect Earth-like planets with MAC-style weathering to exhibit. Without monumental advances in observational techniques \citep[e.g.][]{turyshev2020direct}, direct observation of these parameters on exoplanets is not feasible. Even on Earth, where observations are many orders of magnitude more dense and sensitive than they ever could be for an exoplanet, quantitative understanding of the factors controlling weathering is in its early stages. However, progress is still possible through a combination of theory, modeling, generalization from Earth observations where appropriate, and, eventually, ``statistical comparative planetology'' of the sort described by \citet{bean2017, checlair2019statistical}.}

\textcolor{black}{Theory and modeling can be used to deepen understanding of the relationships between the parameters controlling weathering. In this study, we ignored a variety of potentially important feedbacks between variables in the MAC model. For example, a greater erosion rate decreases soil thickness ($\sim L$), increasing the production rate of soil \citep{heimsath2000soil}, leading to a higher fraction of fresh, weatherable minerals in the soil column ($f_w$). The reduction in soil thickness tends to reduce $D_w$ and solute concentration, while the increase in $f_w$ has the opposite effect, and these opposing responses produce a ``humped'' functional dependence of weathering on erosion, where weathering rates increase with increasing erosion up to a point, beyond which weathering decreases (see Fig. S5 in the supplement to \citet{maher2014hydrologic} for calculations showing this effect). In turn, erosion rates depend on precipitation \citep{perron2017climate}, topographic relief \citep{montgomery2002topographic}, and vegetation cover \citep{collins2004modeling}, among other things. Accounting for the feedback between precipitation and erosion would lead to a stronger coupling between weathering and temperature under circumstances where increased erosion increases weathering rates, as seems to be the case on Earth, where erosion and weathering rates are positively correlated \citep[e.g.][]{gaillardet1999global}. This may make extremely warm climates less likely to occur than a random sampling of parameter combinations in the MAC model would suggest, placing tighter constraints on likely pCO$_2$ values for planets with operative weathering feedbacks. Similarly, since tectonic uplift generates topographic relief which accelerates erosion, we would expect planets with greater globally-averaged uplift rates and/or more extreme topography to tend to be cooler, all else equal. Topography and uplift are determined by the tectonic state of a planet, which is probably strongly tied to variables like the mass and age (thermal history) of a planetary system. This suggests that continued development of theory surrounding the tectonics of Earth-like planets \citep[see e.g.][for discussion]{o2012tectonothermal} could point toward observable correlations between HZ climate and bulk parameters like planetary mass or age. Depending on the expected size of the effects, such trends may be observable with OST/HabEx/LUVOIR-generation telescopes, a variant of the weathering feedback detection method discussed in \citet{checlair2019statistical}.
}

\subsubsection{Limitations of This Study}\label{subsubsec:limitations}
An important limitation of this study is the exclusion of seafloor weathering. Oceanic crust alteration is an important sink of CO$_2$ on Earth, and it is possible that this form of weathering is temperature- or pCO$_2$-dependent, which implies that it may act as a negative feedback on Earth's climate in addition to continental silicate weathering \citep[e.g.][]{Brady:1997p3530,coogan2013evidence, coogan2015alteration}. For example, \citet{krissansen2018constraining} found that the inclusion of temperature- and pH-dependent seafloor weathering significantly moderated climate variations in a geological carbon cycle model of Earth's deep past. If seafloor weathering really does act as a negative feedback on climate, then the sensitivity of climate to land fraction, outgassing, and continental surface properties that we found with the MAC formulation of continental weathering should be less extreme. Our results are intended to highlight the different effects of MAC vs. WHAK continental weathering, and correspond to an extreme limit in which seafloor weathering does not act as a stabilizing or destabilizing feedback. Habitability calculations would be improved by an accurate model of seafloor weathering, but this presents a considerable challenge, as there is no clear consensus as to the nature of the seafloor weathering feedback \citep[e.g.][]{CALDEIRA:1995p2285}, so we did not attempt to model this process. We do, however, suggest that the thermodynamic limit on solute concentration may also be relevant for seafloor weathering: ocean bottom water percolating through hydrothermal systems may reach a maximum concentration of solute analogous to $C_{eq}$ for continental weathering, limiting the importance of kinetic controls on dissolution. In that case, unless the flow rate of water through hydrothermal systems is temperature dependent (analogous to the temperature dependence of runoff on continents), there might be limited scope for seafloor weathering to act as a thermostat. This idea could be explored with reactive transport modeling of seafloor hydrothermal systems. Regardless, the potential for alteration of oceanic crust to stabilize planetary climate is an important subject for further study, and it could have a strong impact on the effective outer edge of the habitable zone.

Another seafloor process left out of our habitability assessment is ``reverse weathering'', a process by which authigenic clay formation on the seafloor absorbs some fraction of the cations delivered to the ocean by silicate weathering, preventing the absorbed cations from forming carbonates to sequester CO$_2$ from the atmosphere-ocean system  \citep[e.g.][]{mackenzie1966chemical,dunlea2017cenozoic,isson2018reverse,trower2019precambrian}. At greater global rates of reverse weathering, fewer moles of CO$_2$ are sequestered per unit silicate weathering, forcing a planet to warm up compared to the no-reverse-weathering case so that enough cations are delivered to the ocean to form carbonates and balance a given amount of outgassing. \citet{isson2018reverse} recently proposed that reverse weathering may provide a stabilizing climate feedback, operating as a complement to the silicate weathering feedback throughout the Precambrian on Earth. If reverse weathering accelerates under high pH (low pCO$_2$) conditions, then its warming effect should become more pronounced when CO$_2$ decreases and less pronounced when CO$_2$ increases, increasing the stability of climate to changes in outgassing (and land fraction) \citep{isson2018reverse}. However, the exact functional form of the pH-dependence of reverse weathering reactions is not currently well-constrained, and the significance of the process as a player in Earth's past climate may well have been moderate \citep{krissansen2020coupled}.

Another shortcoming of our study is the abstraction of inherently three-dimensional processes like precipitation and runoff on continents into zero-dimensional parameterizations. This was appropriate for a first attempt at understanding the implications of the MAC formulation for planetary weathering behavior, but examination of our conclusions with higher dimensional models  would be worthwhile. Previous three-dimensional general circulation model (GCM) global weathering studies have used the WHAK formulation \citep[e.g.][]{donnadieu2006geoclim,edson2012carbonate,paradise2017,paradisehabitable}. Expanding beyond our zero dimensional models would allow us to quantify the importance of things like continental configuration \citep[e.g.][]{lewis2018influence}, atmospheric circulation \citep[e.g.][]{komacek2019atmospheric}, clouds \citep{komacek2019atmospheric}, and rotation rate \citep[e.g.][]{yang2014,jansen2019climates} for planetary precipitation, runoff, and weathering behavior. Importantly, a full GCM study would naturally provide self-consistent treatments of the scalings of runoff and the energetic limit on precipitation with land fraction.

Related to the previous point, this study ignored the crucial and poorly-constrained impacts of clouds on rocky planet climate. Clouds can have a profound impact on planetary albedo: water clouds on Earth reflect away approximately 47.5 W m$^{-2}$ of insolation \citep{stephens2012update}, and simulations of tidally locked planets have found that thick cloud decks at the substellar point can reflect away enough starlight to double the instellation of the inner edge of the HZ for these planets \citep{yang2013}. We implicitly included a cloud albedo effect by setting the surface albedo of our simulations to $a=0.3$, approximately equal to Earth's present-day albedo with clouds, but cloud albedo is intimately linked to atmospheric circulation, so a constant cloud albedo is a grave simplification. Clouds can also have a strong greenhouse effect: water clouds on Earth reduce OLR by approximately 26.4 W m$^{-2}$ \citep{stephens2012update}. However, the particle size distribution for water clouds on exoplanets is completely unconstrained and has a strong impact on their net radiative effect \citep{komacek2019atmospheric}. In addition to water clouds, CO$_2$ ice clouds might be present in the atmospheres of planets with thick CO$_2$ atmospheres \citep[e.g.][]{Forget:1997p3442}. CO$_2$ ice clouds have been found to have strong impacts on albedo and outgoing longwave radiation through their scattering properties in the visible and and near-infrared \citep{Forget:1997p3442,pierrehumbert1998scattering}, although the magnitude of their effect may have been overestimated \citep{kitzmann2016revisiting}.

\subsection{Conclusions}
In this study, we applied the weathering framework developed by \citet{maher2014hydrologic} and extended by \citet{winnick2018relationships} to evaluate rocky planet climate stability across a range of instellations and parameter choices. The MAC weathering framework includes a thermodynamic limit on weathering product concentration in runoff ($C_{eq}$) that previous formulations based on WHAK have not considered. $C_{eq}$ is controlled by pCO$_2$ through the thermodynamics of coupled silicate dissolution and clay precipitation. We also included an energetic limit on global precipitation because instellation drives precipitation through generation of evaporation, meaning the total latent heat flux from global precipitation should not exceed globally averaged instellation (though precipitation may locally exceed absorbed surface instellation slightly due to turbulent heat fluxes from atmosphere to surface, an effect we ignored; see e.g. \citet{Pierrehumbert-2002:hydrologic,o2008hydrological}). 

We found that the MAC formulation leads to runoff- and pCO$_2$-controlled weathering on rocky planets, which leads to interesting changes in planetary weathering behavior compared to simulations with the WHAK formulation that are governed by the kinetics of silicate dissolution. Simulations using the MAC formulation have climates that are more sensitive to CO$_2$ outgassing rate and land fraction, but they are much less sensitive to the details of silicate dissolution kinetics like temperature- and pCO$_2$-dependence. These differences are due to the control of weathering rates in the MAC formulation by the thermodynamic balance between silicate dissolution and clay precipitation instead of the kinetics of silicate dissolution. In the WHAK formulation, increasing temperature increases both runoff and silicate dissolution rate, which increases the ability of WHAK planets to modulate their weathering flux with changes in temperature. In the MAC formulation, temperature only has an impact on weathering through changes in runoff (as well as a mild dependence of $C_{eq}$ on temperature, which we excluded from our calculations), which leads to a weaker temperature-dependence of weathering compared to the exponential dependence displayed by WHAK. This means that decreases in land fraction or increases in volcanic outgassing require larger compensatory temperature changes for planets governed by MAC weathering than for WHAK planets.

We also found that planets governed by MAC-style weathering have climates sensitive to the parameterization of hydrology and surface properties like soil age and soil porosity. Changes to these parameters that make global weathering more effective at delivering cations to the ocean at a given temperature and pCO$_2$ lead to cooler equilibrium climates and move the effective outer edge of the habitable zone closer to the star. The apparent sensitivity of equilibrium rocky planet climate to these parameters suggests that ``Earth-like'' planets may be sensitive to the shifts in surface properties that are likely on a tectonically active surface, e.g. changes in uplift or outgassing rate. There may be significant risk of catastrophic transitions to moist greenhouse or snowball states through stochastic changes to the parameters that control weathering.

Lastly, we showed that the energetic limit on precipitation set by planetary instellation can unintuitively lead to increases in planetary temperature with decreases in instellation. This is because planets that have reached their maximum precipitation at a given instellation lose the feedback on climate provided by modulation of weathering flux in response to temperature through changes in planetary precipitation. Simulations beyond the instellation where the energetic limit kicks in experience a linear decrease in runoff with reduced instellation, which forces pCO$_2$ to increase to higher levels so that $C_{eq}$ can increase to levels that allow runoff to carry enough solute for weathering to balance outgassing. This larger increase in pCO$_2$ with decrease in instellation in the energetically-limited regime leads to the increase in temperature with reduced instellation in some regions of parameter space, though at sufficiently high partial pressures the increase in pCO$_2$ can also lead to cooling as the increased Rayleigh scattering of a thicker atmosphere begins to outweigh the increased greenhouse effect.

In summary, the inclusion of energetic and thermodynamic limits in our simulations of continental silicate weathering on rocky, ocean-bearing exoplanets leads to a diversity of stable climates throughout the habitable zone, ranging from $>$350 K to freezing, depending on a complex interplay of poorly-understood factors. This diversity has important implications for the potential of future telescope missions to infer the operation of the silicate weathering feedback in the habitable zone with population statistics (see Section \ref{subsubsec:observations}). The use of the MAC formulation allows for the explicit incorporation of variables like lithology, hydrology, and soil properties into models of terrestrial planet weathering, which allows for the investigation of a wealth of interesting questions about planetary habitability and the coupling of atmospheres and crusts. Future studies of rocky planet climate and weathering should consider applying the MAC framework instead of, or in addition to, the WHAK framework. 


\acknowledgements{\textbf{Acknowledgements:} We thank Dorian Abbot, Edwin Kite, Tim Lichtenberg, Ramses Ramirez, and an anonymous reviewer for insightful comments and suggestions that improved various versions of this manuscript. RJG thanks the participants at the Rocky Worlds Workshop in Cambridge and the Born\"{o} School on Climate and Conditions for Life on Early Earth and Other Planets for thoughtful discussions on some of the issues covered in this study. RJG acknowledges scholarship funding from the Clarendon Fund and Jesus College, Oxford. RTP is supported by European Research Council Advanced Grant EXOCONDENSE ($\#$740963).}
\bibliographystyle{./aasjournal.bst}
\bibliography{./biblio2.bib}

\end{document}